\documentclass[prl,twocolumn,superscriptaddress,nofootinbib]{revtex4-2}

\usepackage{multirow}%
\usepackage[T1]{fontenc}
\usepackage{xcolor}
\usepackage{pdfcolmk}
\usepackage{mathrsfs}
\usepackage{amsmath}
\usepackage{amssymb}
\usepackage{graphicx}
\usepackage{wasysym}
\PassOptionsToPackage{normalem}{ulem}
\usepackage{ulem}
\usepackage[
unicode=true,pdfusetitle,
bookmarks=true,bookmarksnumbered=false,bookmarksopen=false,
breaklinks=false,pdfborder={0 0 0},pdfborderstyle={},
backref=false,colorlinks=true,
hypertexnames=false
]{hyperref}
\hypersetup{citecolor=blue,urlcolor=blue}

\makeatletter

\providecolor{lyxadded}{rgb}{0,0,1}
\providecolor{lyxdeleted}{rgb}{1,0,0}

\DeclareRobustCommand{\lyxsout}[1]{\ifx\\#1\else\sout{#1}\fi}

\usepackage{bbm}
\usepackage[final]{changes}
\usepackage{ulem}

\begin{document}

\title{Landscape geometry of Majorana zero modes in inhomogeneous superconductors}
\author{Guo-Jian Qiao}
\affiliation{Graduate School of China Academy of Engineering Physics, Beijing 100193, China}
\author{Zhi-Lei Zhang}
\affiliation{Graduate School of China Academy of Engineering Physics, Beijing 100193, China}
\author{Kang Xu}
\affiliation{Graduate School of China Academy of Engineering Physics, Beijing 100193, China}
\author{C. P. Sun}
\email{suncp@gscaep.ac.cn}
\affiliation{Graduate School of China Academy of Engineering Physics, Beijing 100193, China}
 
\begin{abstract}
Conventional methods for characterizing Majorana zero modes in inhomogeneous superconductors generally cannot be applied directly when the potential profile is unknown. To address this limitation, we propose a real-space approach based on a landscape for the zero-energy spatial dynamics. The landscape geometry determines the number of Majorana zero modes, and the conventional topological phase transition is recovered as a geometric transition of the landscape in the homogeneous limit. Without specifying detailed spatial profiles, the approach yields analytical bounds on Hamiltonian parameters and their spatial gradients that guarantee Majorana zero modes. Application to nanowire--superconductor systems demonstrates the predictive power of this profile-independent criterion and quantifies tolerable electrostatic inhomogeneity for Majorana-device design. Landscape geometry thus offers a dynamical perspective with broad applicability to topological boundary states. 
\end{abstract}

\maketitle

\textit{Introduction}---Majorana zero modes are characteristic boundary states of topological superconductors~\cite{Read2000,A_Yu_Kitaev_2001,Fu_2008_SC_Insulator,Lutchyn_2010,Oreg_2010,Alicea_2012,Tanaka2024,review2026}. In homogeneous systems, their existence is characterized by bulk topological invariants defined from the Bloch Hamiltonian~\cite{A_Yu_Kitaev_2001,Chiu_2016}. Realistic semiconductor nanowire--superconductor devices, however, exhibit spatial variations due to disorder, impurities, gate-induced electrostatic potentials, and interface nonuniformity~\cite{Mourik_2012,deng2012anomalous,Das2012_Zero_bias_peaks,Brouwer2012,Marra2017,Loss2018,Woods2021,das2023search,kouwenhoven2025perspective}. Once translational symmetry is lost, momentum is no longer a good quantum number, and the conventional Bloch-band characterization no longer applies directly.

Real-space methods based on Lyapunov exponents~\cite{David2001,DeGottardi2013,Shem2014,Graf2018}, reflection matrices~\cite{Beenakker2011,Fulga2011,Das2016,Wimmer2025SciPost}, and other diagnostics~\cite{Wimmer2014,Loring2017,Lin2021,Marra_2022,Jezequel2022,Pan2024,Jorge2024,Pan2026,Robert2025,ahmed2026} have extended topological characterization to spatially inhomogeneous superconductors. For an actual device, their direct application usually requires specifying its spatial potential profile and then numerically evaluating transfer or scattering matrices, solving eigenvalue equations, etc. In practice, however, the detailed potential profile inside a device is generally unknown. Designing Majorana devices under such uncertainty requires quantitative estimates for tolerable spatial inhomogeneity. This raises a practical question: can one determine in advance bounds on Hamiltonian parameters and their spatial variation that guarantee the existence of Majorana zero modes, without knowing the detailed potential profile?

\begin{figure}
\includegraphics[width=8.5cm]{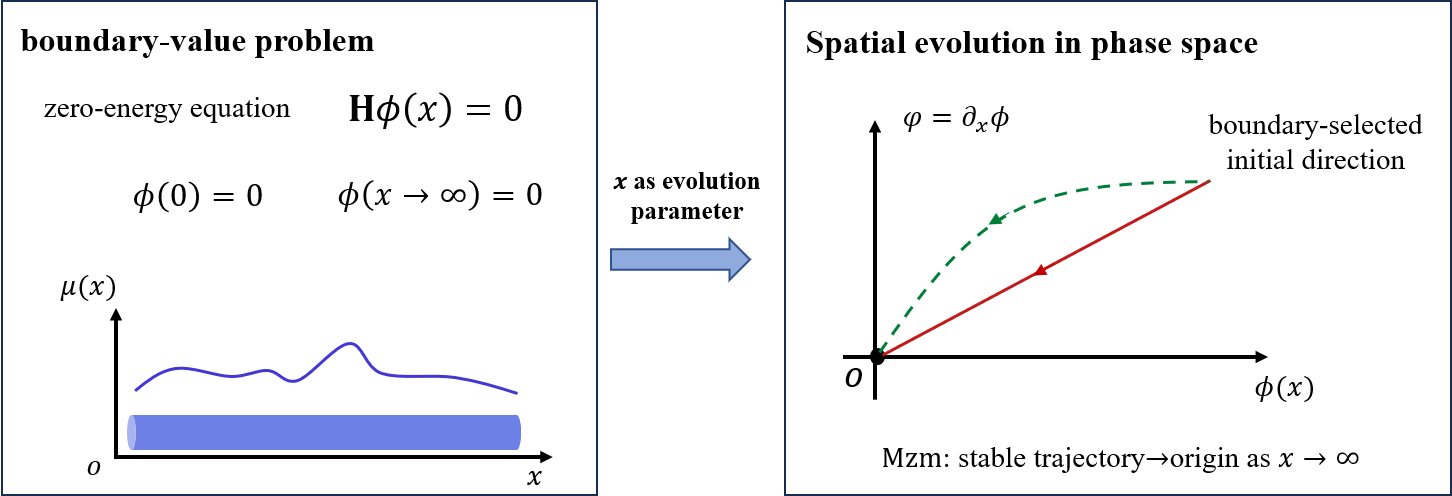}
\caption{A zero-energy equation with spatially inhomogeneous parameters $\mu(x)$ is recast as a spatial dynamical system. The physical boundary condition selects allowed initial directions in phase space, while a Majorana zero mode (Mzm) localized at the left end corresponds to a stable trajectory approaching the origin as $x\rightarrow\infty$.}
\label{Fig1}
\end{figure}

To address this question, we develop a real-space landscape approach for Majorana zero modes in inhomogeneous systems. A Majorana zero mode is determined by a zero-energy equation subject to physical boundary conditions, i.e., a boundary-value problem. Since the detailed potential profile is unknown, we do not solve the equation explicitly but recast it as a first-order dynamical system in which the spatial coordinate $x$ serves as the evolution parameter. The wave function and its spatial derivative then form the phase-space coordinates. A Majorana mode localized at the left end corresponds to a trajectory that starts on the physical boundary at $x=0$ and approaches the origin as $x\rightarrow\infty$, as shown in Fig.~\ref{Fig1}. The problem is thus converted from finding a particular eigenfunction to determining whether the stable subspace of the spatial dynamics intersects the physical boundary subspace. This geometric picture motivates us to introduce a landscape to characterize spatial dynamics; the landscape concept has previously been introduced in other physical settings \cite{Lyapunov1992,pnas_Ao,Wang2008,xu2026wave}.

\begin{figure}
\includegraphics[width=8.3cm]{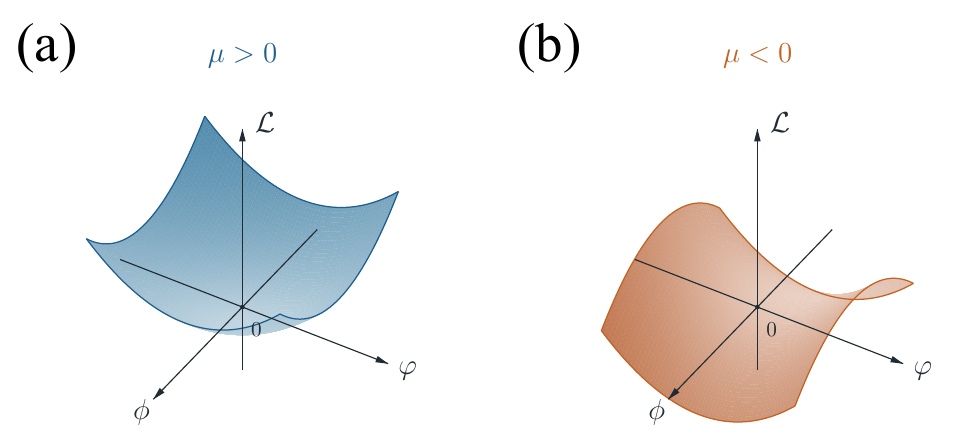}
\caption{Landscape geometry across the topological phase transition. Left: For $\mu>0$, the landscape is positive definite and has a bowl-like geometry around the origin, supporting Majorana zero modes. For $\mu<0$, the landscape is indefinite, and the origin has a saddle geometry with stable and unstable directions, where Majorana zero modes are absent. The topological phase transition, characterized by the presence or absence of Majorana zero modes, is thus manifested as a qualitative change in the landscape geometry.}
\label{Fig2}
\end{figure}

To illustrate this approach, we first construct a landscape for a spinless one-dimensional $p$-wave superconductor with a spatially varying chemical potential. The geometry of the landscape yields a sufficient condition for Majorana zero modes and, in the homogeneous limit, recovers the conventional distinction between topological and trivial phases. The topological phase transition thus appears as a geometric transition of the landscape (see Fig.~\ref{Fig2}).

We then establish a general landscape theorem for zero-energy
boundary states. The landscape geometry fixes the dimensions
of the stable and unstable subspaces, while the zero-mode
count equals the dimension of the intersection between the
stable subspace and the physical boundary subspace. Explicit landscape constructions turn this relation into
analytical bounds on Hamiltonian parameters and their
spatial gradients. These bounds guarantee the existence
of zero-energy boundary states without specifying the
detailed spatial dependence of these parameters or
solving the zero-energy equation.
The dynamical assumptions required by the theorem are also
verified self-consistently using a strict Lyapunov lemma. 

Applied to spatially inhomogeneous semiconductor
nanowire--superconductor hybrid systems, this approach yields
an explicit sufficient bound on the chemical potential
and its spatial gradient that guarantees Majorana
zero modes. Numerical calculations for distinct periodic and quasiperiodic potentials satisfying the same amplitude and gradient bounds demonstrate the predictive power of this profile-independent
criterion. The resulting amplitude--gradient bound
quantifies tolerable electrostatic inhomogeneity
for Majorana-device design. For hybrid systems of InAs--Al and InSb--Al, the maximum width of the local chemical-potential window is of order $1\,\mathrm{meV}$, while the maximum spatial gradient bound is of order $10^{-2}\,\mathrm{meV/nm}$. These estimates offer quantitative guidance for gate-voltage tuning and assessing residual electrostatic nonuniformity.

\textit{Landscape for p-wave superconductors}---We first consider a one-dimensional spinless \(p\)-wave superconductor. Its continuum model is  \cite{Read2000,Brouwer2011prl,marra2025topological} $H=(1/2)\int_{0}^{\infty}\boldsymbol{\psi}^{\dagger}(x)\cdot\mathbf{H}\cdot\boldsymbol{\psi}(x)\mathrm{d} x$,
where $\boldsymbol{\psi}(x)\equiv[\psi(x),\psi^{\dagger}(x)]^{T}$ and the Bogoliubov--de Gennes (BdG) Hamiltonian reads
\begin{equation}
\mathbf{H}=\left[-\frac{\hbar^{2}}{2m}\partial_{x}^{2}-\mu(x)\right]\tau_{z}-i\Delta \tau_{y}\partial_{x}.
\end{equation}
Here, $\mu(x)$ is the spatially dependent chemical potential, $m$ the effective mass, $\Delta$ the $p$-wave pairing strength, and $\tau_i$ the Pauli matrices acting in particle-hole space. Starting from the BdG equation $\mathbf{H}\boldsymbol{\Psi}_n(x)=E_n\boldsymbol{\Psi}_n(x)$, the Hamiltonian can be diagonalized in terms of quasiparticle operators $\gamma_n$. A Majorana quasi-particle is defined by $\gamma_n^\dagger=\gamma_n$. Due to particle-hole symmetry, such a quasiparticle corresponds to a zero-energy state \cite{Alicea_2012,Li_2014,Qiao_2022,Qiao2024,Yue2026,zhang2025poor,zhang2026sensitive}. By decomposing the zero-energy wave function into its real and imaginary parts, the BdG equation becomes
\begin{equation}
	\left[-\frac{\hbar^2}{2m}\partial_x^2-\mu(x)-\Delta \partial_x \right] \phi(x)=0,
\label{eq:realeq}
\end{equation}
with a second branch obtained by \(\Delta \rightarrow - \Delta\). We
first focus on Eq.~\eqref{eq:realeq}. 

Define $\boldsymbol{\Phi}(x)=[\phi(x),\varphi(x)]^{T}$ with $\varphi(x)=\partial_x\phi(x)$, so that the second-order zero-mode equation becomes a first-order spatial dynamical system, 
\begin{equation}
\partial_{x}\mathbf{\Phi}(x)=\mathbf{M}(x)\mathbf{\Phi}(x),\;  \mathbf{M}(x) = \begin{bmatrix}
0 & 1 \\
- \frac{2m\mu(x)}{\hbar^{2}} & - \frac{2m\Delta}{\hbar^{2}}
\end{bmatrix}.
\end{equation}
It is worth noting that \(x\) plays the role of an evolution parameter, while $\boldsymbol{\Phi}(x)$ specifies the state in phase space and traces a trajectory as $x$ evolves. The open boundary at the left end imposes \(\phi(0) = 0\), so admissible initial conditions lie on a one-dimensional boundary line in phase space. Localization further requires \(\mathbf{\Phi}(x\rightarrow\infty)\rightarrow 0\). The existence of a Majorana zero mode localized at the left end therefore becomes the dynamical question: does the physical boundary line contain an initial condition whose spatial trajectory decays to the origin as $x\rightarrow\infty$?

To address this question without explicitly solving the trajectory, we introduce the quadratic landscape
\begin{equation}
\mathcal{L}(x)=\frac{1}{2}\frac{\hbar^{2}}{2m\mu(x)}\varphi^{2}(x)+\frac{1}{2}\phi^{2}(x).
\end{equation}
This quadratic form defines the local geometry of phase space. For $\mu(x)>0$, $\mathcal{L}$ is positive definite and its level sets are closed ellipses around the origin; for $\mu(x)<0$, it becomes indefinite and its level sets acquire a saddle geometry. Differentiating along the trajectory generated by Eq.~\eqref{eq:realeq}, the cross terms cancel exactly, giving
\begin{equation}
\partial_x\mathcal{L}
=-\frac{1}{\mu(x)}
\left[
\Delta+\frac{\hbar^{2}}{4m}\partial_x\ln\mu(x)
\right]\varphi^{2}(x).
\label{eq5}
\end{equation}
Hence, for $\mu(x)>0$ and $\partial_x\ln\mu(x)>-4m\Delta/\hbar^{2}$, the landscape is nonincreasing with $x$. Because $\mathcal{L}$ is positive definite, the trajectory remains bounded, and the zero-energy dynamics admits a trajectory that approaches the origin as $x\rightarrow\infty$. For the right-end Majorana zero mode, we reverse the spatial
coordinate and take the left end infinitely far away.
The same landscape analysis then gives the opposite
gradient constraint. Combining the two bounds yields a sufficient condition
for Majorana zero modes at both ends: 
\begin{equation}
\mu(x)>0,\qquad
\left|\partial_x\ln\mu(x)\right|<
\frac{4m\Delta}{\hbar^{2}},
\label{eq6}
\end{equation}
for all $x\geq0$. The coordinate-reversal argument and a self-consistent verification that Eq.~\eqref{eq6} guarantees the existence of Majorana zero modes, together with the required regularity and uniformity conditions, are all given in Sec.~\ref{sec:sm-pwave} of the Supplemental Material.

Equation~\eqref{eq6} reveals the key physical insight of our approach. In a uniform system, the existence of Majorana zero modes is determined by whether the chemical potential lies in the topological regime \cite{Read2000,Alicea_2012}. In an inhomogeneous system, an additional constraint emerges on the spatial variation: the chemical potential may vary, provided that its local value and spatial rate of variation remain within the sufficient bounds. Thus, perfect spatial uniformity is not required for the existence of Majorana zero modes: sufficiently weak and smooth inhomogeneity can preserve the localized zero-energy states.

In the homogeneous limit, \(\mu(x) = \mu\), Eq. \eqref{eq5} reduces to \(\partial_{x}\mathcal{L}= -(\Delta/\mu)\varphi^{2}\). For $\mu>0$, the landscape is positive definite and monotonically decreasing, and Majorana zero modes exist. For $\mu<0$, the origin has a saddle geometry and no Majorana zero mode can form (see Fig.~\ref{Fig2}). The geometry of the landscape distinguishes the topological and trivial phases \cite{A_Yu_Kitaev_2001,Alicea_2012,Chiu_2016}. The familiar topological transition is thus manifested as a qualitative change in the landscape geometry, providing a real-space counterpart of the conventional Bloch-band description of boundary states. 

\emph{Landscape geometry of zero modes}---We now formulate the approach more generally for zero-energy states in one-dimensional (or quasi-one-dimensional) systems whose eigenvalue equations can be recast as first-order spatial evolution equations, 
\begin{equation}
\partial_{x}\mathbf{\Phi}(x)=\mathbf{M}(x)\mathbf{\Phi}(x),\quad \mathbf{\Phi} \in \mathbb{C}^{2n}.
\label{eq7}
\end{equation}
Let \(\mathbf{G}(x)\) be a continuously differentiable Hermitian matrix that is uniformly bounded and uniformly nonsingular. Define the quadratic landscape $\mathcal{L}(x,\mathbf{\Phi})=(1/2)\mathbf{\Phi}^{\dagger}\mathbf{G}(x)\mathbf{\Phi}$. The spatial-evolution equation then gives
\begin{equation}
\partial_x\mathcal{L}
=\frac{1}{2}\mathbf{\Phi}^{\dagger}\mathbf{\Gamma}(x)\mathbf{\Phi},
\end{equation}
where $\mathbf{\Gamma}(x)=\partial_x\mathbf{G}(x)+\mathbf{G}(x)\mathbf{M}(x)+\mathbf{M}^{\dagger}(x)\mathbf{G}(x)$. When $\mathbf{\Gamma}(x)\leq0$, $\mathcal{L}$ is nonincreasing along every spatial trajectory. Let the inertia of \(\mathbf{G}\) be 
\begin{equation}
\mathrm{In}\lbrack \mathbf{G}\rbrack = \left( n_{pos},n_{neg},0 \right),\quad\quad n_{pos} + n_{neg} = 2n.
\end{equation}

Assume that the initial-value space admits an exponential stable--unstable splitting $E^s\oplus E^u$. In this splitting, initial conditions in $E^s$ generate trajectories that approach the origin exponentially as $x\to\infty$, whereas those in $E^u$ generate trajectories that grow exponentially under spatial evolution, as specified in Eq.~\eqref{app:anchored-splitting} of Sec.~\ref{sec:sm-theorem} in the Supplemental Material. Such a splitting is ensured, for example, by an exponential dichotomy~\cite{coppel1978dichotomies,sacker1978spectral}. Theorem~1 in Sec.~\ref{sec:sm-theorem} then gives
\begin{equation}
\dim E^{s}=n_{pos},\qquad
\dim E^{u}=n_{neg}.
\end{equation}
Thus, the positive and negative inertia indices of the landscape determine the numbers of independent decaying and growing initial directions, respectively.

\begin{figure}
\includegraphics[width=8.5cm]{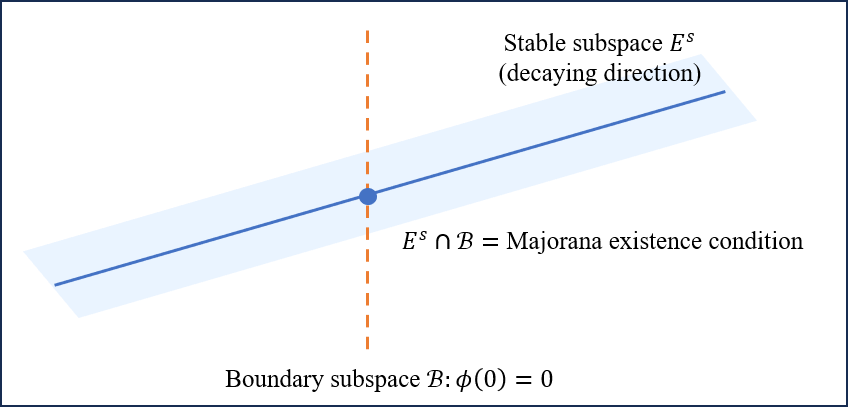}
\caption{Schematic illustration of the stable subspace and boundary conditions for Majorana zero modes. The landscape inertia determines the dimension of the stable subspace $E^s$ (light-blue region), while the boundary subspace $\mathcal B$ contains the physically allowed initial conditions (light-yellow dashed line). Majorana zero modes correspond to nontrivial intersections $E^s\cap\mathcal B$.}
\label{Fig3}
\end{figure}

A decaying solution alone, however, does not constitute a physical zero-energy boundary state, such as a Majorana zero mode. The open boundary imposes additional constraints on the wave function, which define the physical boundary subspace $\mathcal B$. The number of zero modes is therefore given by 
\begin{equation}
\mathcal{N} =\dim\left(E^s\cap\mathcal B\right)= n_{pos}-\mathrm{rank} \mathbf{A}_{s},
\label{eq11}
\end{equation}
where \(\mathbf{A}_{s}\) projects the stable subspace onto the boundary wave-function components, as illustrated schematically in Fig. \ref{Fig3}. For $n_{\mathrm{pos}}\geq n$, a transverse intersection gives
$\operatorname{rank}\mathbf A_s=n$, and therefore
\begin{equation}
\mathcal{N }= n_{pos}-n=\frac{n_{pos} - n_{neg}}{2}.
\label{eq12}
\end{equation}

Equation~\eqref{eq11} implies $\mathcal N\ge n_{\mathrm{pos}}-n$, since $\mathrm{rank} \mathbf{A}_{s} \le n$. Under the stability assumptions above, $n_{\mathrm{pos}}>n$ therefore guarantees at least one boundary zero mode without assuming transversality or determining the individual decaying solutions. Equation~\eqref{eq12} gives the exact zero-mode count under the stated transversality condition, whereas topological protection is determined by the symmetry class. Particle-hole symmetry alone protects only the parity of the Majorana zero-mode number, while an additional chiral symmetry can support an integer-valued topological index~\cite{Chiu_2016,qiao2025size}. 

Once an explicit landscape is constructed, its inertia and monotonicity conditions yield analytical bounds on Hamiltonian parameters and their spatial gradients. These bounds ensure the existence of Majorana zero modes for any spatial profile satisfying them. For both systems considered here, these same bounds also establish the required exponential splitting under the stated uniform bounds and positive margins. This is proved using auxiliary landscapes and Lemma~1 in Sec.~\ref{sec:sm-theorem} of the Supplemental Material. Thus, the applicability of the approach can be verified self-consistently. 

\emph{Estimating tolerable inhomogeneity}---We now apply this geometric result to an inhomogeneous nanowire--superconductor system. In the Nambu basis, the low-energy BdG Hamiltonian reads~\cite{Beenakker2011, Marra2017,Marra_2022}
\begin{equation}
\mathbf{H}=
\left[-\frac{\hbar^{2} \partial_{x}^{2}}{2m}-\mu(x)-i\alpha\sigma_{y}\partial_{x}+B\sigma_{z}\right]\tau_z
-\Delta\sigma_y \tau_y.
\label{nanowire_BdG}
\end{equation}
Here, $\alpha$ is the spin-orbit coupling, $B$ the Zeeman energy, and $\Delta$ the induced pairing gap, while $\mu(x)$ describes spatial inhomogeneity arising, for example, from disorder, impurities, or gate potentials~\cite{Li_2014,Woods2021,Beenakker2011,Brouwer2011,Marra2017,Loss2018}. For Majorana zero modes, separating the electron
component of the BdG wave function into its real and
imaginary parts yields two independent two-component
equations \cite{Qiao_2022}. Each equation can then be recast in the first-order
form of Eq.~\eqref{eq7}, with the four-component
phase-space vector $\boldsymbol{\Phi}(x)$ formed from
the two-component wave function and its spatial derivative.

\begin{figure}
\includegraphics[width=8.6cm]{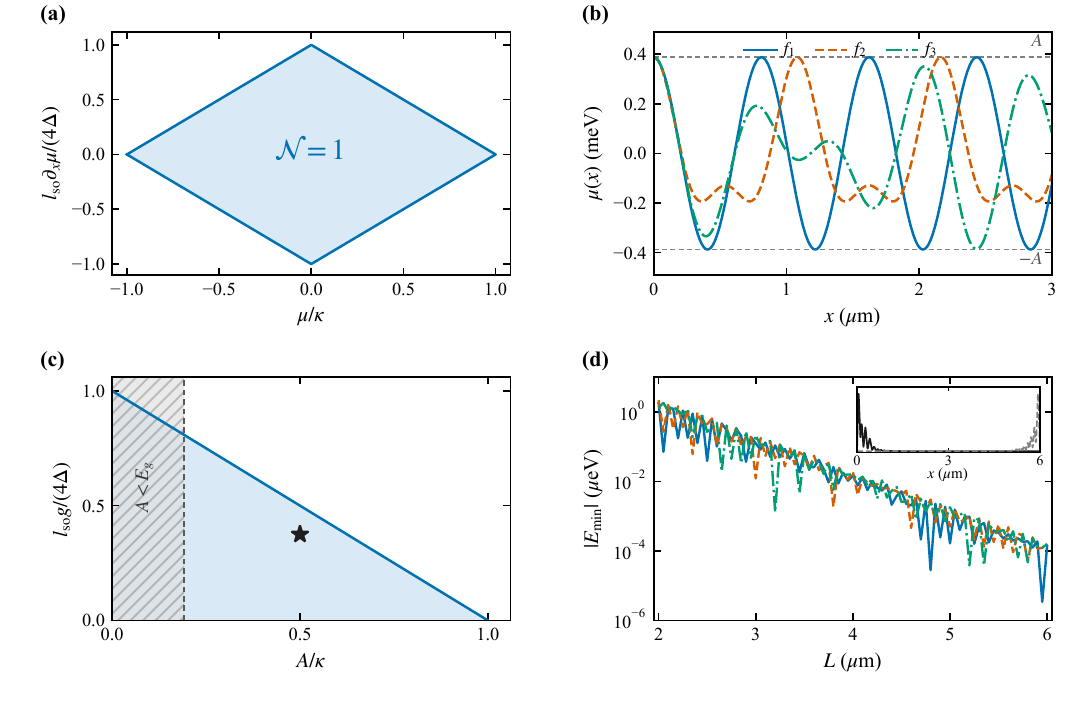}
\caption{(a) The shaded diamond shows the pointwise sufficient region defined by Eq.~\eqref{eq14}, with $\mathcal N=1$ denoting one Majorana zero mode localized at each end of nanowire-superconductor systems. (b) Three potentials $\mu_j(x)=Af_j(q_jx)$ share the bounds $|\mu_j(x)|\leq A=\kappa/2$ and $|\partial_x\mu_j(x)|\leq g=3\,\mu\mathrm{eV/nm}$. Their explicit forms are given in Sec.~\ref{sec:sm-numerics} of the Supplemental Material. (c) Comparison of the landscape criterion $A/\kappa+l_{\mathrm{so}}g/(4\Delta)<1$ (blue) with the gap-preserving norm estimate $A<E_g$ (hatched), where $E_g$ is the bulk gap of the uniform $\mu=0$ reference. The star marks the common bounds in (b). (d) Mean absolute energy of the two eigenstates closest to zero versus wire length; colors and line styles match those in (b). The inset shows the left and right end-mode densities for $f_3$ at $L=6\,\mu\mathrm m$, each rescaled to unit peak. Parameters for (b)--(d): $B=0.8\,\mathrm{meV}$, $\Delta=0.2\,\mathrm{meV}$, $l_{\mathrm{so}}=100\,\mathrm{nm}$, and $m=0.023m_e$.}
\label{Fig4}
\end{figure}

An explicit $4\times4$ landscape matrix is constructed in
Sec.~\ref{sec:sm-nanowire}. For $B>\Delta>0$, combining
the landscape inertia and monotonicity conditions for
the two boundary branches yields the following
sufficient condition:

\begin{equation}
\frac{|\mu(x)|}{\kappa}
+\frac{l_{\mathrm{so}}|\partial_x\mu(x)|}{4\Delta}<1,
\label{eq14}
\end{equation}
where $\kappa=\sqrt{B^{2}-\Delta^{2}}$ and $l_{\mathrm{so}}=\hbar^{2}/(m\alpha)$ is the spin-orbit length~\cite{Nadj2012}. This inequality defines an allowed region in the local chemical potential and its spatial gradient that guarantees the existence of Majorana zero modes localized at both ends [see Fig.~\ref{Fig4}(a)]. The Zeeman energy and induced pairing gap set the maximum
width $2\kappa$ of the local chemical-potential window,
while the pairing gap and spin-orbit length set the
characteristic gradient scale $\Delta/l_{\mathrm{so}}$. The admissible gradient is largest at $\mu=0$ and decreases linearly to zero as $|\mu|$ approaches $\kappa$. These results apply generally to nanowire--superconductor hybrid systems, irrespective of the specific materials. It therefore identifies in advance an allowed amplitude--gradient region for Majorana-device design, without solving the zero-energy equation or even specifying the detailed potential profile. For hybrid systems of InAs and InSb, representative parameters are $l_{\mathrm{so}}\sim100\,\mathrm{nm}$ and $\Delta\sim0.2\,\mathrm{meV}$~\cite{Mourik_2012,Lutchyn2018,Nadj2012}. Taking $B=0.8\,\mathrm{meV}$ gives a maximum width $2\kappa\simeq1.55\,\mathrm{meV}$ for the local chemical-potential window, on the meV scale discussed for hybrid systems~\cite{Qiao2024}. The characteristic gradient scale is $\Delta/l_{\mathrm{so}}\sim2\,\mu\mathrm{eV/nm}$, and the maximum gradient bound is $4\Delta/l_{\mathrm{so}}\sim8\,\mu\mathrm{eV/nm}$ at $\mu=0$. These estimates translate the existence criterion into quantitative tolerances for gate tuning and residual electrostatic nonuniformity.

For continuously differentiable profiles satisfying Eq.~\eqref{eq14} with a uniform positive margin, an auxiliary landscape construction and Lemma~1 establish the exponential splitting required by Theorem~1, as verified in Sec.~\ref{sec:sm-nanowire}. This verifies the applicability of the theorem self-consistently. Equation~\eqref{eq14} therefore guarantees Majorana zero modes
under these conditions. This criterion is sufficient; its violation alone does not
rule out such modes. A complementary criterion for excluding zero-energy boundary states, based on an increasing landscape, is given in Sec.~\ref{sec:sm-theorem}.

To further demonstrate the predictive power of Eq.~\eqref{eq14} when the detailed potential profile is unknown, we characterize the chemical potential solely by a magnitude bound $|\mu(x)|\leq A$ and a gradient bound $|\partial_x\mu(x)|\leq g$. For any continuously differentiable profile satisfying these bounds, the condition $A/\kappa+l_{\mathrm{so}}g/(4\Delta)<1$ guarantees the existence of Majorana zero modes. Figure~\ref{Fig4}(b) shows three distinct periodic and quasiperiodic potentials that remain spatially modulated at arbitrarily large distances. They share the bounds $A\simeq0.387\,\mathrm{meV}$ and $g=3\,\mu\mathrm{eV/nm}$, marked by the star in Fig.~\ref{Fig4}(c). These common bounds give $A/\kappa+l_{\mathrm{so}}g/(4\Delta)\simeq0.875<1$. For comparison, the gray region in Fig.~\ref{Fig4}(c) represents the gap-preserving norm estimate $A<E_g$, where $E_g\simeq0.148\,\mathrm{meV}$ is the bulk gap of the uniform $\mu=0$ reference. The landscape criterion uses both magnitude and gradient
bounds to certify these profiles even when $A>E_g$,
beyond the range covered by this particular norm estimate.  

Figure~\ref{Fig4}(d) shows that the mean magnitude of the two near-zero energies exhibits a decreasing envelope with increasing wire length for all three profiles. The inset shows the probability densities of two states localized at opposite ends of the wire for the quasiperiodic potential at $L=6\,\mu\mathrm m$. The nonzero energies arise from overlap between the end states in a finite wire, while exact zero modes exist in the semi-infinite limit, as guaranteed by the analytical criterion. These results thus demonstrate the profile-independent predictive power of the criterion. The potential profiles, reference-gap estimate, and numerical calculations are detailed in Sec.~\ref{sec:sm-numerics} of the Supplemental Material. An additional analytical application to a localized Gaussian modulation is presented in Sec.~\ref{sec:sm-analysis}.

\emph{Conclusion}---We have developed a real-space landscape approach to characterize Majorana zero modes in spatially inhomogeneous superconductors. By treating the spatial coordinate as an evolution parameter, the zero-energy boundary-value problem is recast as a spatial dynamical problem, in which Majorana zero modes correspond to stable trajectories satisfying the physical boundary conditions. The geometry of the landscape determines the dimensions of the stable and unstable subspaces, while the dimension of the intersection between the stable subspace and the physical boundary subspace gives the number of zero-energy boundary states. These relations establish a general real-space approach for determining the existence and number of such states. In the homogeneous limit, the topological transition emerges as a geometric transition of the landscape, recovering the conventional distinction between topological and trivial phases.

Applied to spinless $p$-wave superconductors and semiconductor nanowire--superconductor hybrid systems, this approach yields analytical sufficient conditions for Majorana zero modes in terms of bounds on Hamiltonian parameters and their spatial gradients. These bounds guarantee the existence of Majorana zero modes for entire classes of spatial profiles satisfying them, without requiring knowledge of their detailed spatial form.  Under the stated uniform bounds and positive margins, auxiliary landscapes and
a strict Lyapunov lemma establish the dynamical conditions required by the counting theorem, completing the
self-consistent justification of these existence bounds. Numerical calculations on finite-length wires with distinct periodic and quasi-periodic potentials demonstrate the predictive power of this profile-independent criterion. For hybird systems of InAs--Al and InSb--Al, the maximum width of the local chemical-potential window is of order $1\,\mathrm{meV}$, while the maximum gradient bound is approximately $8\,\mu\mathrm{eV/nm}$. These estimates translate the existence criterion into quantitative tolerances for gate-voltage tuning and residual electrostatic nonuniformity. The sufficient bounds may be conservative, but they remain useful for designing actual devices, whose detailed internal potential profiles are generally unknown.

Finally, we stress that, while applied here primarily to
Majorana zero modes, the landscape approach characterizes
a general class of zero-energy boundary states whose
eigenstate equations admit a suitable first-order spatial
dynamical formulation. A separate study~\cite{qiao2026landscapeindex} develops an intrinsic real-space topological index from this
geometry for chiral systems and establishes its bulk
and interface correspondences. Through dimensional reduction, that study also extends the index and its boundary correspondence to quantum anomalous Hall systems with periodic and quasiperiodic
modulations of the Dirac mass and to an effective radial chiral model with a hard-wall inner boundary. Landscape geometry thus offers a dynamical perspective on topological boundary states directly in real space, with broad applicability across different systems
and spatial dimensions.

\vspace{0.6em}

\emph{Acknowledgments}---The authors thank Yu Chen at GSCAEP, Sheng-Wen Li at BIT, Xin Yue at CSRC, and Hao-Di Liu and Xin Wu at NENU for helpful discussions. This study is supported by the National Natural Science Foundation of China (Grant Nos. 12688201 and 12547124), the China Postdoctoral Science Foundation (Grant Nos. 2025M784438 and GZC20261845), and the Science Challenge Project (Grant No. TZ2025017).

\vspace{0.6em}
\emph{Data availability}---The data and code used to generate Fig.~\ref{Fig4} are available from the authors upon request. 

\twocolumngrid
\bibliography{main}

\begin{thebibliography}{58}%
\makeatletter
\providecommand \@ifxundefined [1]{%
 \@ifx{#1\undefined}
}%
\providecommand \@ifnum [1]{%
 \ifnum #1\expandafter \@firstoftwo
 \else \expandafter \@secondoftwo
 \fi
}%
\providecommand \@ifx [1]{%
 \ifx #1\expandafter \@firstoftwo
 \else \expandafter \@secondoftwo
 \fi
}%
\providecommand \natexlab [1]{#1}%
\providecommand \enquote  [1]{``#1''}%
\providecommand \bibnamefont  [1]{#1}%
\providecommand \bibfnamefont [1]{#1}%
\providecommand \citenamefont [1]{#1}%
\providecommand \href@noop [0]{\@secondoftwo}%
\providecommand \href [0]{\begingroup \@sanitize@url \@href}%
\providecommand \@href[1]{\@@startlink{#1}\@@href}%
\providecommand \@@href[1]{\endgroup#1\@@endlink}%
\providecommand \@sanitize@url [0]{\catcode `\\12\catcode `\$12\catcode
  `\&12\catcode `\#12\catcode `\^12\catcode `\_12\catcode `\%12\relax}%
\providecommand \@@startlink[1]{}%
\providecommand \@@endlink[0]{}%
\providecommand \url  [0]{\begingroup\@sanitize@url \@url }%
\providecommand \@url [1]{\endgroup\@href {#1}{\urlprefix }}%
\providecommand \urlprefix  [0]{URL }%
\providecommand \Eprint [0]{\href }%
\providecommand \doibase [0]{https://doi.org/}%
\providecommand \selectlanguage [0]{\@gobble}%
\providecommand \bibinfo  [0]{\@secondoftwo}%
\providecommand \bibfield  [0]{\@secondoftwo}%
\providecommand \translation [1]{[#1]}%
\providecommand \BibitemOpen [0]{}%
\providecommand \bibitemStop [0]{}%
\providecommand \bibitemNoStop [0]{.\EOS\space}%
\providecommand \EOS [0]{\spacefactor3000\relax}%
\providecommand \BibitemShut  [1]{\csname bibitem#1\endcsname}%
\let\auto@bib@innerbib\@empty
\bibitem [{\citenamefont {Read}\ and\ \citenamefont {Green}(2000)}]{Read2000}%
  \BibitemOpen
  \bibfield  {author} {\bibinfo {author} {\bibfnamefont {N.}~\bibnamefont
  {Read}}\ and\ \bibinfo {author} {\bibfnamefont {D.}~\bibnamefont {Green}},\
  }\bibfield  {title} {\bibinfo {title} {Paired states of fermions in two
  dimensions with breaking of parity and time-reversal symmetries and the
  fractional quantum hall effect},\ }\href
  {https://doi.org/10.1103/PhysRevB.61.10267} {\bibfield  {journal} {\bibinfo
  {journal} {Phys. Rev. B}\ }\textbf {\bibinfo {volume} {61}},\ \bibinfo
  {pages} {10267} (\bibinfo {year} {2000})}\BibitemShut {NoStop}%
\bibitem [{\citenamefont {Kitaev}(2001)}]{A_Yu_Kitaev_2001}%
  \BibitemOpen
  \bibfield  {author} {\bibinfo {author} {\bibfnamefont {A.~Y.}\ \bibnamefont
  {Kitaev}},\ }\bibfield  {title} {\bibinfo {title} {Unpaired majorana fermions
  in quantum wires},\ }\href {https://doi.org/10.1070/1063-7869/44/10S/S29}
  {\bibfield  {journal} {\bibinfo  {journal} {Physics-Uspekhi}\ }\textbf
  {\bibinfo {volume} {44}},\ \bibinfo {pages} {131} (\bibinfo {year}
  {2001})}\BibitemShut {NoStop}%
\bibitem [{\citenamefont {Fu}\ and\ \citenamefont
  {Kane}(2008)}]{Fu_2008_SC_Insulator}%
  \BibitemOpen
  \bibfield  {author} {\bibinfo {author} {\bibfnamefont {L.}~\bibnamefont
  {Fu}}\ and\ \bibinfo {author} {\bibfnamefont {C.~L.}\ \bibnamefont {Kane}},\
  }\bibfield  {title} {\bibinfo {title} {Superconducting proximity effect and
  majorana fermions at the surface of a topological insulator},\ }\href
  {https://doi.org/10.1103/PhysRevLett.100.096407} {\bibfield  {journal}
  {\bibinfo  {journal} {Phys. Rev. Lett.}\ }\textbf {\bibinfo {volume} {100}},\
  \bibinfo {pages} {096407} (\bibinfo {year} {2008})}\BibitemShut {NoStop}%
\bibitem [{\citenamefont {Lutchyn}\ \emph {et~al.}(2010)\citenamefont
  {Lutchyn}, \citenamefont {Sau},\ and\ \citenamefont
  {Das~Sarma}}]{Lutchyn_2010}%
  \BibitemOpen
  \bibfield  {author} {\bibinfo {author} {\bibfnamefont {R.~M.}\ \bibnamefont
  {Lutchyn}}, \bibinfo {author} {\bibfnamefont {J.~D.}\ \bibnamefont {Sau}},\
  and\ \bibinfo {author} {\bibfnamefont {S.}~\bibnamefont {Das~Sarma}},\
  }\bibfield  {title} {\bibinfo {title} {Majorana fermions and a topological
  phase transition in semiconductor-superconductor heterostructures},\ }\href
  {https://doi.org/10.1103/PhysRevLett.105.077001} {\bibfield  {journal}
  {\bibinfo  {journal} {Phys. Rev. Lett.}\ }\textbf {\bibinfo {volume} {105}},\
  \bibinfo {pages} {077001} (\bibinfo {year} {2010})}\BibitemShut {NoStop}%
\bibitem [{\citenamefont {Oreg}\ \emph {et~al.}(2010)\citenamefont {Oreg},
  \citenamefont {Refael},\ and\ \citenamefont {von Oppen}}]{Oreg_2010}%
  \BibitemOpen
  \bibfield  {author} {\bibinfo {author} {\bibfnamefont {Y.}~\bibnamefont
  {Oreg}}, \bibinfo {author} {\bibfnamefont {G.}~\bibnamefont {Refael}},\ and\
  \bibinfo {author} {\bibfnamefont {F.}~\bibnamefont {von Oppen}},\ }\bibfield
  {title} {\bibinfo {title} {Helical liquids and majorana bound states in
  quantum wires},\ }\href {https://doi.org/10.1103/PhysRevLett.105.177002}
  {\bibfield  {journal} {\bibinfo  {journal} {Phys. Rev. Lett.}\ }\textbf
  {\bibinfo {volume} {105}},\ \bibinfo {pages} {177002} (\bibinfo {year}
  {2010})}\BibitemShut {NoStop}%
\bibitem [{\citenamefont {Alicea}(2012)}]{Alicea_2012}%
  \BibitemOpen
  \bibfield  {author} {\bibinfo {author} {\bibfnamefont {J.}~\bibnamefont
  {Alicea}},\ }\bibfield  {title} {\bibinfo {title} {New directions in the
  pursuit of majorana fermions in solid state systems},\ }\href
  {https://doi.org/10.1088/0034-4885/75/7/076501} {\bibfield  {journal}
  {\bibinfo  {journal} {Reports on Progress in Physics}\ }\textbf {\bibinfo
  {volume} {75}},\ \bibinfo {pages} {076501} (\bibinfo {year}
  {2012})}\BibitemShut {NoStop}%
\bibitem [{\citenamefont {Tanaka}\ \emph {et~al.}(2024)\citenamefont {Tanaka},
  \citenamefont {Tamura},\ and\ \citenamefont {Cayao}}]{Tanaka2024}%
  \BibitemOpen
  \bibfield  {author} {\bibinfo {author} {\bibfnamefont {Y.}~\bibnamefont
  {Tanaka}}, \bibinfo {author} {\bibfnamefont {S.}~\bibnamefont {Tamura}},\
  and\ \bibinfo {author} {\bibfnamefont {J.}~\bibnamefont {Cayao}},\ }\bibfield
   {title} {\bibinfo {title} {Theory of majorana zero modes in unconventional
  superconductors},\ }\href {https://doi.org/10.1093/ptep/ptae065} {\bibfield
  {journal} {\bibinfo  {journal} {Progress of Theoretical and Experimental
  Physics}\ }\textbf {\bibinfo {volume} {2024}},\ \bibinfo {pages} {08C105}
  (\bibinfo {year} {2024})}\BibitemShut {NoStop}%
\bibitem [{\citenamefont {Qiao}\ \emph
  {et~al.}(2026{\natexlab{a}})\citenamefont {Qiao}, \citenamefont {Yue},
  \citenamefont {Zhang},\ and\ \citenamefont {Sun}}]{review2026}%
  \BibitemOpen
  \bibfield  {author} {\bibinfo {author} {\bibfnamefont {G.-J.}\ \bibnamefont
  {Qiao}}, \bibinfo {author} {\bibfnamefont {X.}~\bibnamefont {Yue}}, \bibinfo
  {author} {\bibfnamefont {Z.-L.}\ \bibnamefont {Zhang}},\ and\ \bibinfo
  {author} {\bibfnamefont {C.~P.}\ \bibnamefont {Sun}},\ }\bibfield  {title}
  {\bibinfo {title} {Theoretical study of majorana fermions in hybrid systems
  and experimental observation challenges},\ }\href
  {https://wulixb.iphy.ac.cn/cn/article/doi/10.7498/aps.75.20251512} {\bibfield
   {journal} {\bibinfo  {journal} {Acta Phys. Sin.}\ }\textbf {\bibinfo
  {volume} {75}} (\bibinfo {year} {2026}{\natexlab{a}})}\BibitemShut {NoStop}%
\bibitem [{\citenamefont {Chiu}\ \emph {et~al.}(2016)\citenamefont {Chiu},
  \citenamefont {Teo}, \citenamefont {Schnyder},\ and\ \citenamefont
  {Ryu}}]{Chiu_2016}%
  \BibitemOpen
  \bibfield  {author} {\bibinfo {author} {\bibfnamefont {C.-K.}\ \bibnamefont
  {Chiu}}, \bibinfo {author} {\bibfnamefont {J.~C.~Y.}\ \bibnamefont {Teo}},
  \bibinfo {author} {\bibfnamefont {A.~P.}\ \bibnamefont {Schnyder}},\ and\
  \bibinfo {author} {\bibfnamefont {S.}~\bibnamefont {Ryu}},\ }\bibfield
  {title} {\bibinfo {title} {Classification of topological quantum matter with
  symmetries},\ }\href {https://doi.org/10.1103/RevModPhys.88.035005}
  {\bibfield  {journal} {\bibinfo  {journal} {Rev. Mod. Phys.}\ }\textbf
  {\bibinfo {volume} {88}},\ \bibinfo {pages} {035005} (\bibinfo {year}
  {2016})}\BibitemShut {NoStop}%
\bibitem [{\citenamefont {Mourik}\ \emph {et~al.}(2012)\citenamefont {Mourik},
  \citenamefont {Zuo}, \citenamefont {Frolov}, \citenamefont {Plissard},
  \citenamefont {Bakkers},\ and\ \citenamefont {Kouwenhoven}}]{Mourik_2012}%
  \BibitemOpen
  \bibfield  {author} {\bibinfo {author} {\bibfnamefont {V.}~\bibnamefont
  {Mourik}}, \bibinfo {author} {\bibfnamefont {K.}~\bibnamefont {Zuo}},
  \bibinfo {author} {\bibfnamefont {S.~M.}\ \bibnamefont {Frolov}}, \bibinfo
  {author} {\bibfnamefont {S.~R.}\ \bibnamefont {Plissard}}, \bibinfo {author}
  {\bibfnamefont {E.~P. A.~M.}\ \bibnamefont {Bakkers}},\ and\ \bibinfo
  {author} {\bibfnamefont {L.~P.}\ \bibnamefont {Kouwenhoven}},\ }\bibfield
  {title} {\bibinfo {title} {Signatures of majorana fermions in hybrid
  superconductor-semiconductor nanowire devices},\ }\href
  {https://doi.org/10.1126/science.1222360} {\bibfield  {journal} {\bibinfo
  {journal} {Science}\ }\textbf {\bibinfo {volume} {336}},\ \bibinfo {pages}
  {1003} (\bibinfo {year} {2012})}\BibitemShut {NoStop}%
\bibitem [{\citenamefont {Deng}\ \emph {et~al.}(2012)\citenamefont {Deng},
  \citenamefont {Yu}, \citenamefont {Huang}, \citenamefont {Larsson},
  \citenamefont {Caroff},\ and\ \citenamefont {Xu}}]{deng2012anomalous}%
  \BibitemOpen
  \bibfield  {author} {\bibinfo {author} {\bibfnamefont {M.}~\bibnamefont
  {Deng}}, \bibinfo {author} {\bibfnamefont {C.}~\bibnamefont {Yu}}, \bibinfo
  {author} {\bibfnamefont {G.}~\bibnamefont {Huang}}, \bibinfo {author}
  {\bibfnamefont {M.}~\bibnamefont {Larsson}}, \bibinfo {author} {\bibfnamefont
  {P.}~\bibnamefont {Caroff}},\ and\ \bibinfo {author} {\bibfnamefont
  {H.}~\bibnamefont {Xu}},\ }\bibfield  {title} {\bibinfo {title} {Anomalous
  zero-bias conductance peak in a nb--insb nanowire--nb hybrid device},\ }\href
  {https://pubs.acs.org/doi/10.1021/nl303758w} {\bibfield  {journal} {\bibinfo
  {journal} {Nano letters}\ }\textbf {\bibinfo {volume} {12}},\ \bibinfo
  {pages} {6414} (\bibinfo {year} {2012})}\BibitemShut {NoStop}%
\bibitem [{\citenamefont {Das}\ \emph {et~al.}(2012)\citenamefont {Das},
  \citenamefont {Ronen}, \citenamefont {Most}, \citenamefont {Oreg},
  \citenamefont {Heiblum},\ and\ \citenamefont
  {Shtrikman}}]{Das2012_Zero_bias_peaks}%
  \BibitemOpen
  \bibfield  {author} {\bibinfo {author} {\bibfnamefont {A.}~\bibnamefont
  {Das}}, \bibinfo {author} {\bibfnamefont {Y.}~\bibnamefont {Ronen}}, \bibinfo
  {author} {\bibfnamefont {Y.}~\bibnamefont {Most}}, \bibinfo {author}
  {\bibfnamefont {Y.}~\bibnamefont {Oreg}}, \bibinfo {author} {\bibfnamefont
  {M.}~\bibnamefont {Heiblum}},\ and\ \bibinfo {author} {\bibfnamefont
  {H.}~\bibnamefont {Shtrikman}},\ }\bibfield  {title} {\bibinfo {title}
  {Zero-bias peaks and splitting in an al-inas nanowire topological
  superconductor as a signature of majorana fermions},\ }\href
  {https://doi.org/10.1038/nphys2479} {\bibfield  {journal} {\bibinfo
  {journal} {Nature Physics}\ }\textbf {\bibinfo {volume} {8}},\ \bibinfo
  {pages} {887} (\bibinfo {year} {2012})}\BibitemShut {NoStop}%
\bibitem [{\citenamefont {Kells}\ \emph {et~al.}(2012)\citenamefont {Kells},
  \citenamefont {Meidan},\ and\ \citenamefont {Brouwer}}]{Brouwer2012}%
  \BibitemOpen
  \bibfield  {author} {\bibinfo {author} {\bibfnamefont {G.}~\bibnamefont
  {Kells}}, \bibinfo {author} {\bibfnamefont {D.}~\bibnamefont {Meidan}},\ and\
  \bibinfo {author} {\bibfnamefont {P.~W.}\ \bibnamefont {Brouwer}},\
  }\bibfield  {title} {\bibinfo {title} {Near-zero-energy end states in
  topologically trivial spin-orbit coupled superconducting nanowires with a
  smooth confinement},\ }\href {https://doi.org/10.1103/PhysRevB.86.100503}
  {\bibfield  {journal} {\bibinfo  {journal} {Phys. Rev. B}\ }\textbf {\bibinfo
  {volume} {86}},\ \bibinfo {pages} {100503} (\bibinfo {year}
  {2012})}\BibitemShut {NoStop}%
\bibitem [{\citenamefont {Marra}\ and\ \citenamefont
  {Cuoco}(2017)}]{Marra2017}%
  \BibitemOpen
  \bibfield  {author} {\bibinfo {author} {\bibfnamefont {P.}~\bibnamefont
  {Marra}}\ and\ \bibinfo {author} {\bibfnamefont {M.}~\bibnamefont {Cuoco}},\
  }\bibfield  {title} {\bibinfo {title} {Controlling majorana states in
  topologically inhomogeneous superconductors},\ }\href
  {https://doi.org/10.1103/PhysRevB.95.140504} {\bibfield  {journal} {\bibinfo
  {journal} {Phys. Rev. B}\ }\textbf {\bibinfo {volume} {95}},\ \bibinfo
  {pages} {140504} (\bibinfo {year} {2017})}\BibitemShut {NoStop}%
\bibitem [{\citenamefont {Aseev}\ \emph {et~al.}(2018)\citenamefont {Aseev},
  \citenamefont {Klinovaja},\ and\ \citenamefont {Loss}}]{Loss2018}%
  \BibitemOpen
  \bibfield  {author} {\bibinfo {author} {\bibfnamefont {P.~P.}\ \bibnamefont
  {Aseev}}, \bibinfo {author} {\bibfnamefont {J.}~\bibnamefont {Klinovaja}},\
  and\ \bibinfo {author} {\bibfnamefont {D.}~\bibnamefont {Loss}},\ }\bibfield
  {title} {\bibinfo {title} {Lifetime of majorana qubits in rashba nanowires
  with nonuniform chemical potential},\ }\href
  {https://doi.org/10.1103/PhysRevB.98.155414} {\bibfield  {journal} {\bibinfo
  {journal} {Phys. Rev. B}\ }\textbf {\bibinfo {volume} {98}},\ \bibinfo
  {pages} {155414} (\bibinfo {year} {2018})}\BibitemShut {NoStop}%
\bibitem [{\citenamefont {Woods}\ \emph {et~al.}(2021)\citenamefont {Woods},
  \citenamefont {Das~Sarma},\ and\ \citenamefont {Stanescu}}]{Woods2021}%
  \BibitemOpen
  \bibfield  {author} {\bibinfo {author} {\bibfnamefont {B.~D.}\ \bibnamefont
  {Woods}}, \bibinfo {author} {\bibfnamefont {S.}~\bibnamefont {Das~Sarma}},\
  and\ \bibinfo {author} {\bibfnamefont {T.~D.}\ \bibnamefont {Stanescu}},\
  }\bibfield  {title} {\bibinfo {title} {Charge-impurity effects in hybrid
  majorana nanowires},\ }\href
  {https://doi.org/10.1103/PhysRevApplied.16.054053} {\bibfield  {journal}
  {\bibinfo  {journal} {Phys. Rev. Appl.}\ }\textbf {\bibinfo {volume} {16}},\
  \bibinfo {pages} {054053} (\bibinfo {year} {2021})}\BibitemShut {NoStop}%
\bibitem [{\citenamefont {Das~Sarma}(2023)}]{das2023search}%
  \BibitemOpen
  \bibfield  {author} {\bibinfo {author} {\bibfnamefont {S.}~\bibnamefont
  {Das~Sarma}},\ }\bibfield  {title} {\bibinfo {title} {In search of
  majorana},\ }\href {https://www.nature.com/articles/s41567-022-01900-9}
  {\bibfield  {journal} {\bibinfo  {journal} {Nature Physics}\ }\textbf
  {\bibinfo {volume} {19}},\ \bibinfo {pages} {165} (\bibinfo {year}
  {2023})}\BibitemShut {NoStop}%
\bibitem [{\citenamefont {Kouwenhoven}(2025)}]{kouwenhoven2025perspective}%
  \BibitemOpen
  \bibfield  {author} {\bibinfo {author} {\bibfnamefont {L.}~\bibnamefont
  {Kouwenhoven}},\ }\bibfield  {title} {\bibinfo {title} {Perspective on
  majorana bound-states in hybrid superconductor-semiconductor nanowires},\
  }\href {https://inspirehep.net/literature/2863573} {\bibfield  {journal}
  {\bibinfo  {journal} {Modern Physics Letters B}\ }\textbf {\bibinfo {volume}
  {39}},\ \bibinfo {pages} {2540002} (\bibinfo {year} {2025})}\BibitemShut
  {NoStop}%
\bibitem [{\citenamefont {Motrunich}\ \emph {et~al.}(2001)\citenamefont
  {Motrunich}, \citenamefont {Damle},\ and\ \citenamefont {Huse}}]{David2001}%
  \BibitemOpen
  \bibfield  {author} {\bibinfo {author} {\bibfnamefont {O.}~\bibnamefont
  {Motrunich}}, \bibinfo {author} {\bibfnamefont {K.}~\bibnamefont {Damle}},\
  and\ \bibinfo {author} {\bibfnamefont {D.~A.}\ \bibnamefont {Huse}},\
  }\bibfield  {title} {\bibinfo {title} {Griffiths effects and quantum critical
  points in dirty superconductors without spin-rotation invariance:
  One-dimensional examples},\ }\href
  {https://doi.org/10.1103/PhysRevB.63.224204} {\bibfield  {journal} {\bibinfo
  {journal} {Phys. Rev. B}\ }\textbf {\bibinfo {volume} {63}},\ \bibinfo
  {pages} {224204} (\bibinfo {year} {2001})}\BibitemShut {NoStop}%
\bibitem [{\citenamefont {DeGottardi}\ \emph {et~al.}(2013)\citenamefont
  {DeGottardi}, \citenamefont {Sen},\ and\ \citenamefont
  {Vishveshwara}}]{DeGottardi2013}%
  \BibitemOpen
  \bibfield  {author} {\bibinfo {author} {\bibfnamefont {W.}~\bibnamefont
  {DeGottardi}}, \bibinfo {author} {\bibfnamefont {D.}~\bibnamefont {Sen}},\
  and\ \bibinfo {author} {\bibfnamefont {S.}~\bibnamefont {Vishveshwara}},\
  }\bibfield  {title} {\bibinfo {title} {Majorana fermions in superconducting
  1d systems having periodic, quasiperiodic, and disordered potentials},\
  }\href {https://doi.org/10.1103/PhysRevLett.110.146404} {\bibfield  {journal}
  {\bibinfo  {journal} {Phys. Rev. Lett.}\ }\textbf {\bibinfo {volume} {110}},\
  \bibinfo {pages} {146404} (\bibinfo {year} {2013})}\BibitemShut {NoStop}%
\bibitem [{\citenamefont {Mondragon-Shem}\ \emph {et~al.}(2014)\citenamefont
  {Mondragon-Shem}, \citenamefont {Hughes}, \citenamefont {Song},\ and\
  \citenamefont {Prodan}}]{Shem2014}%
  \BibitemOpen
  \bibfield  {author} {\bibinfo {author} {\bibfnamefont {I.}~\bibnamefont
  {Mondragon-Shem}}, \bibinfo {author} {\bibfnamefont {T.~L.}\ \bibnamefont
  {Hughes}}, \bibinfo {author} {\bibfnamefont {J.}~\bibnamefont {Song}},\ and\
  \bibinfo {author} {\bibfnamefont {E.}~\bibnamefont {Prodan}},\ }\bibfield
  {title} {\bibinfo {title} {Topological criticality in the chiral-symmetric
  aiii class at strong disorder},\ }\href
  {https://doi.org/10.1103/PhysRevLett.113.046802} {\bibfield  {journal}
  {\bibinfo  {journal} {Phys. Rev. Lett.}\ }\textbf {\bibinfo {volume} {113}},\
  \bibinfo {pages} {046802} (\bibinfo {year} {2014})}\BibitemShut {NoStop}%
\bibitem [{\citenamefont {Graf}\ and\ \citenamefont
  {Shapiro}(2018)}]{Graf2018}%
  \BibitemOpen
  \bibfield  {author} {\bibinfo {author} {\bibfnamefont {G.~M.}\ \bibnamefont
  {Graf}}\ and\ \bibinfo {author} {\bibfnamefont {J.}~\bibnamefont {Shapiro}},\
  }\bibfield  {title} {\bibinfo {title} {The bulk-edge correspondence for
  disordered chiral chains},\ }\href
  {https://doi.org/10.1007/s00220-018-3247-0} {\bibfield  {journal} {\bibinfo
  {journal} {Communications in Mathematical Physics}\ }\textbf {\bibinfo
  {volume} {363}},\ \bibinfo {pages} {829} (\bibinfo {year}
  {2018})}\BibitemShut {NoStop}%
\bibitem [{\citenamefont {Akhmerov}\ \emph {et~al.}(2011)\citenamefont
  {Akhmerov}, \citenamefont {Dahlhaus}, \citenamefont {Hassler}, \citenamefont
  {Wimmer},\ and\ \citenamefont {Beenakker}}]{Beenakker2011}%
  \BibitemOpen
  \bibfield  {author} {\bibinfo {author} {\bibfnamefont {A.~R.}\ \bibnamefont
  {Akhmerov}}, \bibinfo {author} {\bibfnamefont {J.~P.}\ \bibnamefont
  {Dahlhaus}}, \bibinfo {author} {\bibfnamefont {F.}~\bibnamefont {Hassler}},
  \bibinfo {author} {\bibfnamefont {M.}~\bibnamefont {Wimmer}},\ and\ \bibinfo
  {author} {\bibfnamefont {C.~W.~J.}\ \bibnamefont {Beenakker}},\ }\bibfield
  {title} {\bibinfo {title} {Quantized conductance at the majorana phase
  transition in a disordered superconducting wire},\ }\href
  {https://doi.org/10.1103/PhysRevLett.106.057001} {\bibfield  {journal}
  {\bibinfo  {journal} {Phys. Rev. Lett.}\ }\textbf {\bibinfo {volume} {106}},\
  \bibinfo {pages} {057001} (\bibinfo {year} {2011})}\BibitemShut {NoStop}%
\bibitem [{\citenamefont {Fulga}\ \emph {et~al.}(2011)\citenamefont {Fulga},
  \citenamefont {Hassler}, \citenamefont {Akhmerov},\ and\ \citenamefont
  {Beenakker}}]{Fulga2011}%
  \BibitemOpen
  \bibfield  {author} {\bibinfo {author} {\bibfnamefont {I.~C.}\ \bibnamefont
  {Fulga}}, \bibinfo {author} {\bibfnamefont {F.}~\bibnamefont {Hassler}},
  \bibinfo {author} {\bibfnamefont {A.~R.}\ \bibnamefont {Akhmerov}},\ and\
  \bibinfo {author} {\bibfnamefont {C.~W.~J.}\ \bibnamefont {Beenakker}},\
  }\bibfield  {title} {\bibinfo {title} {Scattering formula for the topological
  quantum number of a disordered multimode wire},\ }\href
  {https://doi.org/10.1103/PhysRevB.83.155429} {\bibfield  {journal} {\bibinfo
  {journal} {Phys. Rev. B}\ }\textbf {\bibinfo {volume} {83}},\ \bibinfo
  {pages} {155429} (\bibinfo {year} {2011})}\BibitemShut {NoStop}%
\bibitem [{\citenamefont {Das~Sarma}\ \emph {et~al.}(2016)\citenamefont
  {Das~Sarma}, \citenamefont {Nag},\ and\ \citenamefont {Sau}}]{Das2016}%
  \BibitemOpen
  \bibfield  {author} {\bibinfo {author} {\bibfnamefont {S.}~\bibnamefont
  {Das~Sarma}}, \bibinfo {author} {\bibfnamefont {A.}~\bibnamefont {Nag}},\
  and\ \bibinfo {author} {\bibfnamefont {J.~D.}\ \bibnamefont {Sau}},\
  }\bibfield  {title} {\bibinfo {title} {How to infer non-abelian statistics
  and topological visibility from tunneling conductance properties of realistic
  majorana nanowires},\ }\href {https://doi.org/10.1103/PhysRevB.94.035143}
  {\bibfield  {journal} {\bibinfo  {journal} {Phys. Rev. B}\ }\textbf {\bibinfo
  {volume} {94}},\ \bibinfo {pages} {035143} (\bibinfo {year}
  {2016})}\BibitemShut {NoStop}%
\bibitem [{\citenamefont {Day}\ \emph {et~al.}(2025)\citenamefont {Day},
  \citenamefont {Manesco}, \citenamefont {Wimmer},\ and\ \citenamefont
  {Akhmerov}}]{Wimmer2025SciPost}%
  \BibitemOpen
  \bibfield  {author} {\bibinfo {author} {\bibfnamefont {I.~A.}\ \bibnamefont
  {Day}}, \bibinfo {author} {\bibfnamefont {A.~L.~R.}\ \bibnamefont {Manesco}},
  \bibinfo {author} {\bibfnamefont {M.}~\bibnamefont {Wimmer}},\ and\ \bibinfo
  {author} {\bibfnamefont {A.~R.}\ \bibnamefont {Akhmerov}},\ }\bibfield
  {title} {\bibinfo {title} {{Identifying biases of the Majorana scattering
  invariant}},\ }\href {https://doi.org/10.21468/SciPostPhysCore.8.3.047}
  {\bibfield  {journal} {\bibinfo  {journal} {SciPost Phys. Core}\ }\textbf
  {\bibinfo {volume} {8}},\ \bibinfo {pages} {047} (\bibinfo {year}
  {2025})}\BibitemShut {NoStop}%
\bibitem [{\citenamefont {Adagideli}\ \emph {et~al.}(2014)\citenamefont
  {Adagideli}, \citenamefont {Wimmer},\ and\ \citenamefont
  {Teker}}]{Wimmer2014}%
  \BibitemOpen
  \bibfield  {author} {\bibinfo {author} {\bibfnamefont {i.~d.~I.}\
  \bibnamefont {Adagideli}}, \bibinfo {author} {\bibfnamefont {M.}~\bibnamefont
  {Wimmer}},\ and\ \bibinfo {author} {\bibfnamefont {A.}~\bibnamefont
  {Teker}},\ }\bibfield  {title} {\bibinfo {title} {Effects of electron
  scattering on the topological properties of nanowires: Majorana fermions from
  disorder and superlattices},\ }\href
  {https://doi.org/10.1103/PhysRevB.89.144506} {\bibfield  {journal} {\bibinfo
  {journal} {Phys. Rev. B}\ }\textbf {\bibinfo {volume} {89}},\ \bibinfo
  {pages} {144506} (\bibinfo {year} {2014})}\BibitemShut {NoStop}%
\bibitem [{\citenamefont {Loring}\ and\ \citenamefont
  {Schulz-Baldes}(2017)}]{Loring2017}%
  \BibitemOpen
  \bibfield  {author} {\bibinfo {author} {\bibfnamefont {T.~A.}\ \bibnamefont
  {Loring}}\ and\ \bibinfo {author} {\bibfnamefont {H.}~\bibnamefont
  {Schulz-Baldes}},\ }\bibfield  {title} {\bibinfo {title} {Finite volume
  calculation of {K}-theory invariants},\ }\href
  {https://nyjm.albany.edu/j/2017/23_1111.html} {\bibfield  {journal} {\bibinfo
   {journal} {New York Journal of Mathematics}\ }\textbf {\bibinfo {volume}
  {23}},\ \bibinfo {pages} {1111} (\bibinfo {year} {2017})}\BibitemShut
  {NoStop}%
\bibitem [{\citenamefont {Lin}\ \emph {et~al.}(2021)\citenamefont {Lin},
  \citenamefont {Ke},\ and\ \citenamefont {Lee}}]{Lin2021}%
  \BibitemOpen
  \bibfield  {author} {\bibinfo {author} {\bibfnamefont {L.}~\bibnamefont
  {Lin}}, \bibinfo {author} {\bibfnamefont {Y.}~\bibnamefont {Ke}},\ and\
  \bibinfo {author} {\bibfnamefont {C.}~\bibnamefont {Lee}},\ }\bibfield
  {title} {\bibinfo {title} {Real-space representation of the winding number
  for a one-dimensional chiral-symmetric topological insulator},\ }\href
  {https://doi.org/10.1103/PhysRevB.103.224208} {\bibfield  {journal} {\bibinfo
   {journal} {Phys. Rev. B}\ }\textbf {\bibinfo {volume} {103}},\ \bibinfo
  {pages} {224208} (\bibinfo {year} {2021})}\BibitemShut {NoStop}%
\bibitem [{\citenamefont {Marra}\ and\ \citenamefont
  {Nigro}(2022)}]{Marra_2022}%
  \BibitemOpen
  \bibfield  {author} {\bibinfo {author} {\bibfnamefont {P.}~\bibnamefont
  {Marra}}\ and\ \bibinfo {author} {\bibfnamefont {A.}~\bibnamefont {Nigro}},\
  }\bibfield  {title} {\bibinfo {title} {Majorana/andreev crossover and the
  fate of the topological phase transition in inhomogeneous nanowires},\ }\href
  {https://doi.org/10.1088/1361-648X/ac44d2} {\bibfield  {journal} {\bibinfo
  {journal} {Journal of Physics: Condensed Matter}\ }\textbf {\bibinfo {volume}
  {34}},\ \bibinfo {pages} {124001} (\bibinfo {year} {2022})}\BibitemShut
  {NoStop}%
\bibitem [{\citenamefont {Jezequel}\ \emph {et~al.}(2022)\citenamefont
  {Jezequel}, \citenamefont {Tauber},\ and\ \citenamefont
  {Delplace}}]{Jezequel2022}%
  \BibitemOpen
  \bibfield  {author} {\bibinfo {author} {\bibfnamefont {L.}~\bibnamefont
  {Jezequel}}, \bibinfo {author} {\bibfnamefont {C.}~\bibnamefont {Tauber}},\
  and\ \bibinfo {author} {\bibfnamefont {P.}~\bibnamefont {Delplace}},\
  }\bibfield  {title} {\bibinfo {title} {Estimating bulk and edge topological
  indices in finite open chiral chains},\ }\href
  {https://doi.org/10.1063/5.0096720} {\bibfield  {journal} {\bibinfo
  {journal} {Journal of Mathematical Physics}\ }\textbf {\bibinfo {volume}
  {63}},\ \bibinfo {pages} {121901} (\bibinfo {year} {2022})}\BibitemShut
  {NoStop}%
\bibitem [{\citenamefont {Pan}\ and\ \citenamefont
  {Das~Sarma}(2024)}]{Pan2024}%
  \BibitemOpen
  \bibfield  {author} {\bibinfo {author} {\bibfnamefont {H.}~\bibnamefont
  {Pan}}\ and\ \bibinfo {author} {\bibfnamefont {S.}~\bibnamefont
  {Das~Sarma}},\ }\bibfield  {title} {\bibinfo {title} {Disordered majorana
  nanowires: Studying disorder without any disorder},\ }\href
  {https://doi.org/10.1103/PhysRevB.110.075401} {\bibfield  {journal} {\bibinfo
   {journal} {Phys. Rev. B}\ }\textbf {\bibinfo {volume} {110}},\ \bibinfo
  {pages} {075401} (\bibinfo {year} {2024})}\BibitemShut {NoStop}%
\bibitem [{\citenamefont {Awoga}\ and\ \citenamefont
  {Cayao}(2024)}]{Jorge2024}%
  \BibitemOpen
  \bibfield  {author} {\bibinfo {author} {\bibfnamefont {O.~A.}\ \bibnamefont
  {Awoga}}\ and\ \bibinfo {author} {\bibfnamefont {J.}~\bibnamefont {Cayao}},\
  }\bibfield  {title} {\bibinfo {title} {Identifying trivial and majorana
  zero-energy modes using the majorana polarization},\ }\href
  {https://doi.org/10.1103/PhysRevB.110.165404} {\bibfield  {journal} {\bibinfo
   {journal} {Phys. Rev. B}\ }\textbf {\bibinfo {volume} {110}},\ \bibinfo
  {pages} {165404} (\bibinfo {year} {2024})}\BibitemShut {NoStop}%
\bibitem [{\citenamefont {Pan}\ and\ \citenamefont
  {Das~Sarma}(2026)}]{Pan2026}%
  \BibitemOpen
  \bibfield  {author} {\bibinfo {author} {\bibfnamefont {H.}~\bibnamefont
  {Pan}}\ and\ \bibinfo {author} {\bibfnamefont {S.}~\bibnamefont
  {Das~Sarma}},\ }\bibfield  {title} {\bibinfo {title} {Majorana zero modes in
  semiconductor-superconductor hybrid structures: Defining topology in short
  and disordered nanowires through majorana splitting},\ }\href
  {https://doi.org/10.1103/2v41-yvs1} {\bibfield  {journal} {\bibinfo
  {journal} {Phys. Rev. B}\ }\textbf {\bibinfo {volume} {113}},\ \bibinfo
  {pages} {165420} (\bibinfo {year} {2026})}\BibitemShut {NoStop}%
\bibitem [{\citenamefont {Eissele}\ \emph {et~al.}(2025)\citenamefont
  {Eissele}, \citenamefont {Roy}, \citenamefont {Tewari},\ and\ \citenamefont
  {Stanescu}}]{Robert2025}%
  \BibitemOpen
  \bibfield  {author} {\bibinfo {author} {\bibfnamefont {R.}~\bibnamefont
  {Eissele}}, \bibinfo {author} {\bibfnamefont {B.~B.}\ \bibnamefont {Roy}},
  \bibinfo {author} {\bibfnamefont {S.}~\bibnamefont {Tewari}},\ and\ \bibinfo
  {author} {\bibfnamefont {T.~D.}\ \bibnamefont {Stanescu}},\ }\bibfield
  {title} {\bibinfo {title} {Topological invariant for finite systems in the
  presence of disorder},\ }\href {https://arxiv.org/abs/2508.13146} {\bibfield
  {journal} {\bibinfo  {journal} {arXiv preprint arXiv:2508.13146}\ } (\bibinfo
  {year} {2025})}\BibitemShut {NoStop}%
\bibitem [{\citenamefont {Ahmed}\ \emph {et~al.}(2026)\citenamefont {Ahmed},
  \citenamefont {Cayao},\ and\ \citenamefont {Tanaka}}]{ahmed2026}%
  \BibitemOpen
  \bibfield  {author} {\bibinfo {author} {\bibfnamefont {E.}~\bibnamefont
  {Ahmed}}, \bibinfo {author} {\bibfnamefont {J.}~\bibnamefont {Cayao}},\ and\
  \bibinfo {author} {\bibfnamefont {Y.}~\bibnamefont {Tanaka}},\ }\href
  {https://arxiv.org/abs/2608.09758} {\bibinfo {title} {Disorder-robust trivial
  majorana-like states from smooth confinement in chiral superconducting
  nanowires}} (\bibinfo {year} {2026}),\ \Eprint
  {https://arxiv.org/abs/2608.09758} {arXiv:2608.09758} \BibitemShut {NoStop}%
\bibitem [{\citenamefont {LYAPUNOV}(1992)}]{Lyapunov1992}%
  \BibitemOpen
  \bibfield  {author} {\bibinfo {author} {\bibfnamefont {A.~M.}\ \bibnamefont
  {LYAPUNOV}},\ }\bibfield  {title} {\bibinfo {title} {The general problem of
  the stability of motion},\ }\href {https://doi.org/10.1080/00207179208934253}
  {\bibfield  {journal} {\bibinfo  {journal} {International Journal of
  Control}\ }\textbf {\bibinfo {volume} {55}},\ \bibinfo {pages} {531}
  (\bibinfo {year} {1992})}\BibitemShut {NoStop}%
\bibitem [{\citenamefont {Kwon}\ \emph {et~al.}(2005)\citenamefont {Kwon},
  \citenamefont {Ao},\ and\ \citenamefont {Thouless}}]{pnas_Ao}%
  \BibitemOpen
  \bibfield  {author} {\bibinfo {author} {\bibfnamefont {C.}~\bibnamefont
  {Kwon}}, \bibinfo {author} {\bibfnamefont {P.}~\bibnamefont {Ao}},\ and\
  \bibinfo {author} {\bibfnamefont {D.~J.}\ \bibnamefont {Thouless}},\
  }\bibfield  {title} {\bibinfo {title} {Structure of stochastic dynamics near
  fixed points},\ }\href {https://doi.org/10.1073/pnas.0506347102} {\bibfield
  {journal} {\bibinfo  {journal} {Proceedings of the National Academy of
  Sciences}\ }\textbf {\bibinfo {volume} {102}},\ \bibinfo {pages} {13029}
  (\bibinfo {year} {2005})}\BibitemShut {NoStop}%
\bibitem [{\citenamefont {Wang}\ \emph {et~al.}(2008)\citenamefont {Wang},
  \citenamefont {Xu},\ and\ \citenamefont {Wang}}]{Wang2008}%
  \BibitemOpen
  \bibfield  {author} {\bibinfo {author} {\bibfnamefont {J.}~\bibnamefont
  {Wang}}, \bibinfo {author} {\bibfnamefont {L.}~\bibnamefont {Xu}},\ and\
  \bibinfo {author} {\bibfnamefont {E.}~\bibnamefont {Wang}},\ }\bibfield
  {title} {\bibinfo {title} {Potential landscape and flux framework of
  nonequilibrium networks: Robustness, dissipation, and coherence of
  biochemical oscillations},\ }\href {https://doi.org/10.1073/pnas.0800579105}
  {\bibfield  {journal} {\bibinfo  {journal} {Proceedings of the National
  Academy of Sciences}\ }\textbf {\bibinfo {volume} {105}},\ \bibinfo {pages}
  {12271} (\bibinfo {year} {2008})}\BibitemShut {NoStop}%
\bibitem [{\citenamefont {Xu}\ \emph {et~al.}(2026)\citenamefont {Xu},
  \citenamefont {Yi}, \citenamefont {Yan},\ and\ \citenamefont
  {Sun}}]{xu2026wave}%
  \BibitemOpen
  \bibfield  {author} {\bibinfo {author} {\bibfnamefont {K.}~\bibnamefont
  {Xu}}, \bibinfo {author} {\bibfnamefont {M.-M.}\ \bibnamefont {Yi}}, \bibinfo
  {author} {\bibfnamefont {Z.-H.}\ \bibnamefont {Yan}},\ and\ \bibinfo {author}
  {\bibfnamefont {C.~P.}\ \bibnamefont {Sun}},\ }\href
  {https://arxiv.org/abs/2605.15658} {\bibinfo {title} {Wave packet landscape
  in open quantum systems}} (\bibinfo {year} {2026}),\ \Eprint
  {https://arxiv.org/abs/2605.15658} {arXiv:2605.15658} \BibitemShut {NoStop}%
\bibitem [{\citenamefont {Brouwer}\ \emph
  {et~al.}(2011{\natexlab{a}})\citenamefont {Brouwer}, \citenamefont
  {Duckheim}, \citenamefont {Romito},\ and\ \citenamefont {von
  Oppen}}]{Brouwer2011prl}%
  \BibitemOpen
  \bibfield  {author} {\bibinfo {author} {\bibfnamefont {P.~W.}\ \bibnamefont
  {Brouwer}}, \bibinfo {author} {\bibfnamefont {M.}~\bibnamefont {Duckheim}},
  \bibinfo {author} {\bibfnamefont {A.}~\bibnamefont {Romito}},\ and\ \bibinfo
  {author} {\bibfnamefont {F.}~\bibnamefont {von Oppen}},\ }\bibfield  {title}
  {\bibinfo {title} {Probability distribution of majorana end-state energies in
  disordered wires},\ }\href {https://doi.org/10.1103/PhysRevLett.107.196804}
  {\bibfield  {journal} {\bibinfo  {journal} {Phys. Rev. Lett.}\ }\textbf
  {\bibinfo {volume} {107}},\ \bibinfo {pages} {196804} (\bibinfo {year}
  {2011}{\natexlab{a}})}\BibitemShut {NoStop}%
\bibitem [{\citenamefont {Marra}\ and\ \citenamefont
  {Nigro}(2025)}]{marra2025topological}%
  \BibitemOpen
  \bibfield  {author} {\bibinfo {author} {\bibfnamefont {P.}~\bibnamefont
  {Marra}}\ and\ \bibinfo {author} {\bibfnamefont {A.}~\bibnamefont {Nigro}},\
  }\bibfield  {title} {\bibinfo {title} {Topological zero modes and bounded
  modes at smooth domain walls: Exact solutions and dualities},\ }\href
  {https://academic.oup.com/ptep/article/2025/2/023A01/7920790?login=false}
  {\bibfield  {journal} {\bibinfo  {journal} {Progress of Theoretical and
  Experimental Physics}\ }\textbf {\bibinfo {volume} {2025}},\ \bibinfo {pages}
  {023A01} (\bibinfo {year} {2025})}\BibitemShut {NoStop}%
\bibitem [{\citenamefont {Li}\ \emph {et~al.}(2014)\citenamefont {Li},
  \citenamefont {Li}, \citenamefont {Cai},\ and\ \citenamefont
  {Sun}}]{Li_2014}%
  \BibitemOpen
  \bibfield  {author} {\bibinfo {author} {\bibfnamefont {S.-W.}\ \bibnamefont
  {Li}}, \bibinfo {author} {\bibfnamefont {Z.-Z.}\ \bibnamefont {Li}}, \bibinfo
  {author} {\bibfnamefont {C.~Y.}\ \bibnamefont {Cai}},\ and\ \bibinfo {author}
  {\bibfnamefont {C.~P.}\ \bibnamefont {Sun}},\ }\bibfield  {title} {\bibinfo
  {title} {Probing zero modes of a defect in a kitaev quantum wire},\ }\href
  {https://doi.org/10.1103/PhysRevB.89.134505} {\bibfield  {journal} {\bibinfo
  {journal} {Phys. Rev. B}\ }\textbf {\bibinfo {volume} {89}},\ \bibinfo
  {pages} {134505} (\bibinfo {year} {2014})}\BibitemShut {NoStop}%
\bibitem [{\citenamefont {Qiao}\ \emph {et~al.}(2022)\citenamefont {Qiao},
  \citenamefont {Li},\ and\ \citenamefont {Sun}}]{Qiao_2022}%
  \BibitemOpen
  \bibfield  {author} {\bibinfo {author} {\bibfnamefont {G.-J.}\ \bibnamefont
  {Qiao}}, \bibinfo {author} {\bibfnamefont {S.-W.}\ \bibnamefont {Li}},\ and\
  \bibinfo {author} {\bibfnamefont {C.~P.}\ \bibnamefont {Sun}},\ }\bibfield
  {title} {\bibinfo {title} {Magnetic field constraint for majorana zero modes
  in a hybrid nanowire},\ }\href {https://doi.org/10.1103/PhysRevB.106.104517}
  {\bibfield  {journal} {\bibinfo  {journal} {Phys. Rev. B}\ }\textbf {\bibinfo
  {volume} {106}},\ \bibinfo {pages} {104517} (\bibinfo {year}
  {2022})}\BibitemShut {NoStop}%
\bibitem [{\citenamefont {Qiao}\ \emph {et~al.}(2024)\citenamefont {Qiao},
  \citenamefont {Yue},\ and\ \citenamefont {Sun}}]{Qiao2024}%
  \BibitemOpen
  \bibfield  {author} {\bibinfo {author} {\bibfnamefont {G.-J.}\ \bibnamefont
  {Qiao}}, \bibinfo {author} {\bibfnamefont {X.}~\bibnamefont {Yue}},\ and\
  \bibinfo {author} {\bibfnamefont {C.~P.}\ \bibnamefont {Sun}},\ }\bibfield
  {title} {\bibinfo {title} {Dressed majorana fermion in a hybrid nanowire},\
  }\href {https://link.aps.org/doi/10.1103/PhysRevLett.133.266605} {\bibfield
  {journal} {\bibinfo  {journal} {Phys. Rev. Lett.}\ }\textbf {\bibinfo
  {volume} {133}},\ \bibinfo {pages} {266605} (\bibinfo {year}
  {2024})}\BibitemShut {NoStop}%
\bibitem [{\citenamefont {Yue}\ \emph {et~al.}(2026)\citenamefont {Yue},
  \citenamefont {Qiao},\ and\ \citenamefont {Sun}}]{Yue2026}%
  \BibitemOpen
  \bibfield  {author} {\bibinfo {author} {\bibfnamefont {X.}~\bibnamefont
  {Yue}}, \bibinfo {author} {\bibfnamefont {G.-J.}\ \bibnamefont {Qiao}},\ and\
  \bibinfo {author} {\bibfnamefont {C.~P.}\ \bibnamefont {Sun}},\ }\bibfield
  {title} {\bibinfo {title} {Finite-size effects on metallization versus chiral
  majorana fermions},\ }\href {https://doi.org/10.1103/57w4-rs6c} {\bibfield
  {journal} {\bibinfo  {journal} {Phys. Rev. B}\ }\textbf {\bibinfo {volume}
  {113}},\ \bibinfo {pages} {115416} (\bibinfo {year} {2026})}\BibitemShut
  {NoStop}%
\bibitem [{\citenamefont {Zhang}\ \emph
  {et~al.}(2026{\natexlab{a}})\citenamefont {Zhang}, \citenamefont {Qiao},\
  and\ \citenamefont {Sun}}]{zhang2025poor}%
  \BibitemOpen
  \bibfield  {author} {\bibinfo {author} {\bibfnamefont {Z.-L.}\ \bibnamefont
  {Zhang}}, \bibinfo {author} {\bibfnamefont {G.-J.}\ \bibnamefont {Qiao}},\
  and\ \bibinfo {author} {\bibfnamefont {C.~P.}\ \bibnamefont {Sun}},\
  }\bibfield  {title} {\bibinfo {title} {Poor man's majorana modes in two
  quantum dots dressed by superconducting quasiexcitations},\ }\href
  {https://doi.org/10.1103/qp5g-942h} {\bibfield  {journal} {\bibinfo
  {journal} {Phys. Rev. B}\ }\textbf {\bibinfo {volume} {113}},\ \bibinfo
  {pages} {195416} (\bibinfo {year} {2026}{\natexlab{a}})}\BibitemShut
  {NoStop}%
\bibitem [{\citenamefont {Zhang}\ \emph
  {et~al.}(2026{\natexlab{b}})\citenamefont {Zhang}, \citenamefont {Yue},
  \citenamefont {Qiao},\ and\ \citenamefont {Sun}}]{zhang2026sensitive}%
  \BibitemOpen
  \bibfield  {author} {\bibinfo {author} {\bibfnamefont {Z.-L.}\ \bibnamefont
  {Zhang}}, \bibinfo {author} {\bibfnamefont {X.}~\bibnamefont {Yue}}, \bibinfo
  {author} {\bibfnamefont {G.-J.}\ \bibnamefont {Qiao}},\ and\ \bibinfo
  {author} {\bibfnamefont {C.~P.}\ \bibnamefont {Sun}},\ }\bibfield  {title}
  {\bibinfo {title} {Sensitive dependence of poor man's majorana modes on the
  length of superconductor},\ }\href {https://arxiv.org/html/2604.12950v1}
  {\bibfield  {journal} {\bibinfo  {journal} {arXiv preprint arXiv:2604.12950}\
  } (\bibinfo {year} {2026}{\natexlab{b}})}\BibitemShut {NoStop}%
\bibitem [{\citenamefont {Coppel}(1978)}]{coppel1978dichotomies}%
  \BibitemOpen
  \bibfield  {author} {\bibinfo {author} {\bibfnamefont {W.~A.}\ \bibnamefont
  {Coppel}},\ }\href {https://doi.org/10.1007/BFb0067780} {\emph {\bibinfo
  {title} {Dichotomies in Stability Theory}}},\ \bibinfo {series} {Lecture
  Notes in Mathematics}, Vol.\ \bibinfo {volume} {629}\ (\bibinfo  {publisher}
  {Springer},\ \bibinfo {address} {Berlin},\ \bibinfo {year}
  {1978})\BibitemShut {NoStop}%
\bibitem [{\citenamefont {Sacker}\ and\ \citenamefont
  {Sell}(1978)}]{sacker1978spectral}%
  \BibitemOpen
  \bibfield  {author} {\bibinfo {author} {\bibfnamefont {R.~J.}\ \bibnamefont
  {Sacker}}\ and\ \bibinfo {author} {\bibfnamefont {G.~R.}\ \bibnamefont
  {Sell}},\ }\bibfield  {title} {\bibinfo {title} {A spectral theory for linear
  differential systems},\ }\href {https://doi.org/10.1016/0022-0396(78)90057-8}
  {\bibfield  {journal} {\bibinfo  {journal} {Journal of Differential
  Equations}\ }\textbf {\bibinfo {volume} {27}},\ \bibinfo {pages} {320}
  (\bibinfo {year} {1978})}\BibitemShut {NoStop}%
\bibitem [{\citenamefont {Qiao}\ \emph {et~al.}(2025)\citenamefont {Qiao},
  \citenamefont {Zhang}, \citenamefont {Yue},\ and\ \citenamefont
  {Sun}}]{qiao2025size}%
  \BibitemOpen
  \bibfield  {author} {\bibinfo {author} {\bibfnamefont {G.-J.}\ \bibnamefont
  {Qiao}}, \bibinfo {author} {\bibfnamefont {Z.-L.}\ \bibnamefont {Zhang}},
  \bibinfo {author} {\bibfnamefont {X.}~\bibnamefont {Yue}},\ and\ \bibinfo
  {author} {\bibfnamefont {C.~P.}\ \bibnamefont {Sun}},\ }\bibfield  {title}
  {\bibinfo {title} {Size optimization for observeing majorana fermions},\
  }\href {https://arxiv.org/abs/2511.21423} {\bibfield  {journal} {\bibinfo
  {journal} {arXiv preprint arXiv:2511.21423}\ } (\bibinfo {year}
  {2025})}\BibitemShut {NoStop}%
\bibitem [{\citenamefont {Brouwer}\ \emph
  {et~al.}(2011{\natexlab{b}})\citenamefont {Brouwer}, \citenamefont
  {Duckheim}, \citenamefont {Romito},\ and\ \citenamefont {von
  Oppen}}]{Brouwer2011}%
  \BibitemOpen
  \bibfield  {author} {\bibinfo {author} {\bibfnamefont {P.~W.}\ \bibnamefont
  {Brouwer}}, \bibinfo {author} {\bibfnamefont {M.}~\bibnamefont {Duckheim}},
  \bibinfo {author} {\bibfnamefont {A.}~\bibnamefont {Romito}},\ and\ \bibinfo
  {author} {\bibfnamefont {F.}~\bibnamefont {von Oppen}},\ }\bibfield  {title}
  {\bibinfo {title} {Topological superconducting phases in disordered quantum
  wires with strong spin-orbit coupling},\ }\href
  {https://doi.org/10.1103/PhysRevB.84.144526} {\bibfield  {journal} {\bibinfo
  {journal} {Phys. Rev. B}\ }\textbf {\bibinfo {volume} {84}},\ \bibinfo
  {pages} {144526} (\bibinfo {year} {2011}{\natexlab{b}})}\BibitemShut
  {NoStop}%
\bibitem [{\citenamefont {Nadj-Perge}\ \emph {et~al.}(2012)\citenamefont
  {Nadj-Perge}, \citenamefont {Pribiag}, \citenamefont {van~den Berg},
  \citenamefont {Zuo}, \citenamefont {Plissard}, \citenamefont {Bakkers},
  \citenamefont {Frolov},\ and\ \citenamefont {Kouwenhoven}}]{Nadj2012}%
  \BibitemOpen
  \bibfield  {author} {\bibinfo {author} {\bibfnamefont {S.}~\bibnamefont
  {Nadj-Perge}}, \bibinfo {author} {\bibfnamefont {V.~S.}\ \bibnamefont
  {Pribiag}}, \bibinfo {author} {\bibfnamefont {J.~W.~G.}\ \bibnamefont
  {van~den Berg}}, \bibinfo {author} {\bibfnamefont {K.}~\bibnamefont {Zuo}},
  \bibinfo {author} {\bibfnamefont {S.~R.}\ \bibnamefont {Plissard}}, \bibinfo
  {author} {\bibfnamefont {E.~P. A.~M.}\ \bibnamefont {Bakkers}}, \bibinfo
  {author} {\bibfnamefont {S.~M.}\ \bibnamefont {Frolov}},\ and\ \bibinfo
  {author} {\bibfnamefont {L.~P.}\ \bibnamefont {Kouwenhoven}},\ }\bibfield
  {title} {\bibinfo {title} {Spectroscopy of spin-orbit quantum bits in indium
  antimonide nanowires},\ }\href
  {https://doi.org/10.1103/PhysRevLett.108.166801} {\bibfield  {journal}
  {\bibinfo  {journal} {Phys. Rev. Lett.}\ }\textbf {\bibinfo {volume} {108}},\
  \bibinfo {pages} {166801} (\bibinfo {year} {2012})}\BibitemShut {NoStop}%
\bibitem [{\citenamefont {Lutchyn}\ \emph {et~al.}(2018)\citenamefont
  {Lutchyn}, \citenamefont {Bakkers}, \citenamefont {Kouwenhoven},
  \citenamefont {Krogstrup}, \citenamefont {Marcus},\ and\ \citenamefont
  {Oreg}}]{Lutchyn2018}%
  \BibitemOpen
  \bibfield  {author} {\bibinfo {author} {\bibfnamefont {R.~M.}\ \bibnamefont
  {Lutchyn}}, \bibinfo {author} {\bibfnamefont {E.~P. A.~M.}\ \bibnamefont
  {Bakkers}}, \bibinfo {author} {\bibfnamefont {L.~P.}\ \bibnamefont
  {Kouwenhoven}}, \bibinfo {author} {\bibfnamefont {P.}~\bibnamefont
  {Krogstrup}}, \bibinfo {author} {\bibfnamefont {C.~M.}\ \bibnamefont
  {Marcus}},\ and\ \bibinfo {author} {\bibfnamefont {Y.}~\bibnamefont {Oreg}},\
  }\bibfield  {title} {\bibinfo {title} {Majorana zero modes in
  superconductor-semiconductor heterostructures},\ }\href
  {https://doi.org/10.1038/s41578-018-0003-1} {\bibfield  {journal} {\bibinfo
  {journal} {Nature Reviews Materials}\ }\textbf {\bibinfo {volume} {3}},\
  \bibinfo {pages} {52} (\bibinfo {year} {2018})}\BibitemShut {NoStop}%
\bibitem [{\citenamefont {Qiao}\ \emph
  {et~al.}(2026{\natexlab{b}})\citenamefont {Qiao}, \citenamefont {Zhang},\
  and\ \citenamefont {Sun}}]{qiao2026landscapeindex}%
  \BibitemOpen
  \bibfield  {author} {\bibinfo {author} {\bibfnamefont {G.-J.}\ \bibnamefont
  {Qiao}}, \bibinfo {author} {\bibfnamefont {Z.-L.}\ \bibnamefont {Zhang}},\
  and\ \bibinfo {author} {\bibfnamefont {C.~P.}\ \bibnamefont {Sun}},\ }\href
  {https://arxiv.org/abs/2611.22500} {\bibinfo {title} {Landscape index and its
  boundary correspondence in inhomogeneous chiral systems}} (\bibinfo {year}
  {2026}{\natexlab{b}}),\ \Eprint {https://arxiv.org/abs/2611.22500}
  {arXiv:2611.22500 [cond-mat.mes-hall]} \BibitemShut {NoStop}%
\bibitem [{\citenamefont {Palmer}(1987)}]{palmer1987perturbation}%
  \BibitemOpen
  \bibfield  {author} {\bibinfo {author} {\bibfnamefont {K.~J.}\ \bibnamefont
  {Palmer}},\ }\bibfield  {title} {\bibinfo {title} {A perturbation theorem for
  exponential dichotomies},\ }\href {https://doi.org/10.1017/S0308210500018175}
  {\bibfield  {journal} {\bibinfo  {journal} {Proceedings of the Royal Society
  of Edinburgh Section A: Mathematics}\ }\textbf {\bibinfo {volume} {106}},\
  \bibinfo {pages} {25} (\bibinfo {year} {1987})}\BibitemShut {NoStop}%
\bibitem [{\citenamefont {Mitropolsky}\ \emph {et~al.}(2002)\citenamefont
  {Mitropolsky}, \citenamefont {Samoilenko},\ and\ \citenamefont
  {Kulik}}]{Mitropolsky2002}%
  \BibitemOpen
  \bibfield  {author} {\bibinfo {author} {\bibfnamefont {Y.~A.}\ \bibnamefont
  {Mitropolsky}}, \bibinfo {author} {\bibfnamefont {A.~M.}\ \bibnamefont
  {Samoilenko}},\ and\ \bibinfo {author} {\bibfnamefont {V.~L.}\ \bibnamefont
  {Kulik}},\ }\href {https://doi.org/10.1201/9781482264890} {\emph {\bibinfo
  {title} {Dichotomies and Stability in Nonautonomous Linear Systems}}},\
  \bibinfo {edition} {1st}\ ed.\ (\bibinfo  {publisher} {CRC Press},\ \bibinfo
  {address} {London},\ \bibinfo {year} {2002})\BibitemShut {NoStop}%
\bibitem [{\citenamefont {Pan}\ and\ \citenamefont
  {Das~Sarma}(2020)}]{Pan2020}%
  \BibitemOpen
  \bibfield  {author} {\bibinfo {author} {\bibfnamefont {H.}~\bibnamefont
  {Pan}}\ and\ \bibinfo {author} {\bibfnamefont {S.}~\bibnamefont
  {Das~Sarma}},\ }\bibfield  {title} {\bibinfo {title} {Physical mechanisms for
  zero-bias conductance peaks in majorana nanowires},\ }\href
  {https://doi.org/10.1103/PhysRevResearch.2.013377} {\bibfield  {journal}
  {\bibinfo  {journal} {Phys. Rev. Res.}\ }\textbf {\bibinfo {volume} {2}},\
  \bibinfo {pages} {013377} (\bibinfo {year} {2020})}\BibitemShut {NoStop}%
\end{thebibliography}%

\onecolumngrid

\setcounter{equation}{0}
\renewcommand{\theequation}{E\arabic{equation}}







\clearpage
\onecolumngrid

\pagenumbering{arabic}
\pagestyle{plain}

\setcounter{section}{0}
\setcounter{secnumdepth}{2}
\renewcommand{\thesection}{S\arabic{section}}

\setcounter{equation}{0}
\renewcommand{\theequation}{S\arabic{equation}}

\setcounter{figure}{0}
\renewcommand{\thefigure}{S\arabic{figure}}
\setcounter{table}{0}
\renewcommand{\thetable}{S\arabic{table}}

\begin{center}
\textbf{Supplemental Material for
``Landscape geometry of Majorana zero modes
in inhomogeneous superconductors''}
\end{center}

\twocolumngrid

\section{Landscape analysis for $P$-wave superconductors}
\label{sec:sm-pwave}

Equation~\eqref{eq:realeq} in the main text describes the zero-energy branch associated with the Majorana mode localized at the left end. The second zero-energy branch is
\begin{equation}
\left\lbrack - \frac{\hbar^{2}}{2m}\partial_{x}^{2} - \mu(x) + \Delta\partial_{x} \right\rbrack\phi_{i}(x) = 0.   
\label{eqA1}
\end{equation}
Hereafter, the subscript $i$ is omitted for simplicity. Introducing the coordinate transformation $y=L-x$, where $L$ denotes the system length, together with $\bar{\phi}(y)=\phi(L-y)$ and $\bar{\mu}(y)=\mu(L-y)$, Eq. \eqref{eqA1} becomes
\begin{equation}
\left\lbrack - \frac{\hbar^{2}}{2m}\partial_{y}^{2} - \bar{\mu}(y) - \Delta\partial_{y} \right\rbrack\bar{\phi}(y) = 0.
\end{equation}
In the semi-infinite limit ($L\rightarrow\infty$), the boundary conditions become $\bar{\phi}(0)=0$ and $\lim_{y\rightarrow\infty}\bar{\phi}(y)=0$. Thus, the equation and boundary conditions are identical to those of the zero-energy equation for the Majorana mode localized at the left end, as considered in the main text. The same landscape analysis therefore gives $\bar{\mu}(y) > 0$ and $\partial_{y}\ln\bar{\mu}(y) > - 4m\Delta/\hbar^{2}$. Returning to the original coordinate with
\(\partial_{y} = - \partial_{x}\) yields
\begin{equation}
\mu(x) > 0,\quad\quad\partial_{x}\ln\mu(x) < \frac{4m\Delta}{\hbar^{2}}.
\end{equation}
Combining the existence conditions for Majorana zero modes localized at the left and right ends yields Eq. \eqref{eq6}. Physically, Majorana zero modes localized at the two ends correspond to stable decaying trajectories along opposite spatial directions, requiring the logarithmic gradient of the chemical potential to satisfy bounds of opposite sign.

We now self-consistently verify that the parameter bounds in Eq.~\eqref{eq6}, together with the regularity and uniformity conditions stated below, ensure that Theorem~1 in Sec.~\ref{sec:sm-theorem} applies. The theorem then guarantees the existence of Majorana zero modes. Assume $\mu\in C^1([0,\infty))$ with
$0<\mu_{\min}\leq\mu(x)\leq\mu_{\max}<\infty$,
and impose Eq.~\eqref{eq6} with a uniform positive margin:
\begin{equation}
|\partial_x\ln\mu(x)|
\leq(1-\eta)\frac{4m\Delta}{\hbar^2},
\qquad 0<\eta<1.
\label{app:pwave-uniform}
\end{equation}
For convenience, introduce the rescaled pairing parameter
$\Delta_s=2m\Delta/\hbar^2$ and chemical potential
$\mu_s(x)=2m\mu(x)/\hbar^2$. In terms of the rescaled parameters, the spatial evolution
matrix and the landscape derivative matrix read
\begin{equation}
\mathbf M(x)=
\begin{bmatrix}
0&1\\
-\mu_s(x)&-\Delta_s
\end{bmatrix},
\qquad
\mathbf\Gamma(x)=
\begin{bmatrix}
0&0\\
0&-r(x)
\end{bmatrix}.
\end{equation}
The rescaled bounds are
$\mu_{s,\min}=2m\mu_{\min}/\hbar^2$ and
$\mu_{s,\max}=2m\mu_{\max}/\hbar^2$.
The uniform gradient bound gives
\begin{equation}
r(x)
=
\frac{2\Delta_s+\mu_s'(x)/\mu_s(x)}{\mu_s(x)}
\geq
\frac{2\Delta_s\eta}{\mu_{s,\max}}
=:r_0>0.
\end{equation}

For the auxiliary landscape matrix
$\widetilde{\mathbf G}(x)=\mathbf G(x)+\epsilon\sigma_x$,
the corresponding derivative matrix is
\begin{equation}
\widetilde{\mathbf\Gamma}(x)=
\begin{bmatrix}
-2\epsilon\mu_s(x)&-\epsilon\Delta_s\\
-\epsilon\Delta_s&-r(x)+2\epsilon
\end{bmatrix}.
\end{equation}
For any phase-space vector $\mathbf\Phi=(\phi,\varphi)^T$,
the weighted Young inequality
\begin{equation}
2\Delta_s|\phi||\varphi|
\leq\mu_{s,\min}|\phi|^2
+\frac{\Delta_s^2}{\mu_{s,\min}}|\varphi|^2
\end{equation}
controls the off-diagonal contribution.
Together with $\mu_s\geq\mu_{s,\min}$ and $r\geq r_0$,
it yields
\begin{equation}
\mathbf\Phi^\dagger\widetilde{\mathbf\Gamma}\mathbf\Phi
\leq-\epsilon\mu_{s,\min}|\phi|^2-(r_0-\epsilon C_p)|\varphi|^2,
\end{equation}
where $C_p=2+\Delta_s^2/\mu_{s,\min}$.

The original landscape satisfies
$\mathbf G\geq g_*\mathbf I_2$, with
$g_*=\min\{1,\mu_{s,\max}^{-1}\}>0$.
Choose a fixed
$0<\epsilon<\min\{g_*,r_0/C_p\}$.
Using $\|\sigma_x\|=1$ and the preceding derivative estimate,
we obtain
\begin{equation}
\widetilde{\mathbf G}(x)
\geq(g_*-\epsilon)\mathbf I_2>0,\quad\quad 
\widetilde{\mathbf\Gamma}(x)
\leq-c_p\mathbf I_2,
\end{equation}
where
$c_p=\min\{\epsilon\mu_{s,\min},r_0-\epsilon C_p\}>0$.

The uniform bounds on $\mu_s$ ensure that $\mathbf M$
and $\widetilde{\mathbf G}$ are uniformly bounded,
while the positive lower bound on $\widetilde{\mathbf G}$
ensures uniform invertibility. All conditions of Lemma~1 in
Sec.~\ref{sec:sm-theorem} are therefore satisfied.
The lemma establishes the exponential splitting in
Eq.~\eqref{app:anchored-splitting},
with $\dim E^s=2$ and $\dim E^u=0$.
This verifies the dynamical condition required to apply
Theorem~1 under the stated uniform bounds and positive margin.
The same verification applies to the right-boundary branch
in the inward coordinate $y$, since the absolute-gradient
bound in Eq.~\eqref{app:pwave-uniform} is invariant under
coordinate reversal. Theorem~1 therefore guarantees the existence of a Majorana zero mode at each end, completing the self-consistency verification of the existence criterion in Eq.~\eqref{eq6}.

\section{General landscape theorem}
\label{sec:sm-theorem}

Consider the first-order differential system
\begin{equation}
\partial_x\mathbf{\Phi}(x)
=
\mathbf{M}(x)\mathbf{\Phi}(x),
\qquad
\mathbf{\Phi}\in\mathbb{C}^{2n}.
\end{equation}
Here, $\mathbf M(x)$ is continuous on $[0,\infty)$, and $U(x,y)$ denotes the spatial evolution operator from $y$ to $x$. We assume that the initial-value space at $x=0$ admits an exponentially separated stable--unstable splitting: there exist linear subspaces $E^s$ and $E^u$, together with constants $K\geq1$ and $\alpha>0$, such that $\mathbb{C}^{2n}=E^s\oplus E^u$ and 
\begin{equation}
\begin{aligned}
&\|U(x,0)v_s\|
\leq
Ke^{-\alpha x}\|v_s\|,
\;\; v_s\in E^s,\\
&\|U(x,0)v_u\|
\geq
K^{-1}e^{\alpha x}\|v_u\|,
\;\; v_u\in E^u.
\end{aligned}
\label{app:anchored-splitting}
\end{equation}

Thus, $E^s$ consists of initial conditions that decay exponentially under forward evolution, while $E^u$ consists of initial conditions that grow exponentially under forward evolution. For every finite $x\geq0$, define $E^s(x)=U(x,0)E^s$ and $E^u(x)=U(x,0)E^u$. Since \(U(x,0)\) is invertible, the decomposition \(\mathbb{C}^{2n}=E^s(x)\oplus E^u(x)\) is preserved, with \(\dim E^s(x)=\dim E^s\) and \(\dim E^u(x)=\dim E^u\). Moreover, for $w\in E^u(x)$, the last inequality in Eq.~\eqref{app:anchored-splitting} is equivalent to
\begin{equation}
\|U(0,x)w\|
\leq
Ke^{-\alpha x}\|w\|.
\label{app:backward-contraction}
\end{equation}

Now define the landscape $\mathcal{L}(x,\mathbf{\Phi})=(1/2)\mathbf{\Phi}^{\dagger}\mathbf{G}(x)\mathbf{\Phi}$, where $\mathbf{G}(x)$ is a continuously differentiable Hermitian matrix with spatially constant inertia
\begin{equation}
\operatorname{In}[\mathbf{G}(x)]
=
(n_{\mathrm{pos}},n_{\mathrm{neg}},0).
\end{equation}
We further assume that $\sup_{x\geq0}\|\mathbf{G}(x)\|<\infty$ (uniformly bounded) and $\sup_{x\geq0}\|\mathbf{G}^{-1}(x)\|<\infty$ (uniformly invertible). If $\mathbf{\Gamma}=\partial_x\mathbf{G}+\mathbf{G}\mathbf{M}+\mathbf{M}^{\dagger}\mathbf{G}\leq0$, every spatial trajectory satisfies $\partial_x\mathcal{L}=(1/2)\mathbf{\Phi}^{\dagger}\mathbf{\Gamma}\mathbf{\Phi}\leq0$.

\textbf{Theorem 1.} Under the assumptions above, the dimensions of the stable and unstable subspaces are determined by the inertia of the landscape:
\begin{equation}
\dim E^s=n_{\mathrm{pos}},
\qquad
\dim E^u=n_{\mathrm{neg}}.
\label{A8}
\end{equation}

\textbf{Proof.} Take any $\Phi_0\in E^s$. By Eq.~\eqref{app:anchored-splitting}, $U(x,0)\Phi_0\rightarrow0$ as $x\rightarrow+\infty$. Since $\mathbf{G}(x)$ is uniformly bounded,
\begin{equation}
\lim_{x\rightarrow\infty}
\mathcal{L}\!\left[x,U(x,0)\Phi_0\right]
=
0.
\end{equation}
Since $\mathcal{L}$ is non-increasing along forward evolution, $\mathcal{L}(0,\Phi_0)\geq0$. Thus, $\mathbf{G}(0)$ is positive semidefinite on $E^s$. To prove $\dim E^s\leq n_{\mathrm{pos}}$, suppose instead that $\dim E^s>n_{\mathrm{pos}}$. Let $H_-(0)$ be the negative spectral subspace of $\mathbf{G}(0)$, with $\dim H_-(0)=n_{\mathrm{neg}}$. The subspace-intersection inequality gives
\begin{align}
\dim\!\left[E^s\cap H_-(0)\right]
&\geq
\dim E^s+\dim H_-(0)-2n
\nonumber\\
&=
\dim E^s-n_{\mathrm{pos}}
>0. \label{E10}
\end{align}

Hence there exists $0\neq v\in E^s\cap H_-(0)$. However, $v\in E^s$ gives $v^{\dagger}\mathbf{G}(0)v\geq0$, whereas $v\in H_-(0)$ gives $v^{\dagger}\mathbf{G}(0)v<0$, which is a contradiction. Therefore, $\dim E^s\leq n_{\mathrm{pos}}$.

We next prove $\dim E^u\leq n_{\mathrm{neg}}$. Suppose, to the contrary, that $\dim E^u>n_{\mathrm{neg}}$. Let $H_+(x)$ denote the positive spectral subspace of $\mathbf{G}(x)$ so that $\dim H_+(x)=n_{\mathrm{pos}}$. For every finite $x\geq0$, $\dim E^u(x)=\dim E^u$, and hence
\begin{align}
\dim\!\left[E^u(x)\cap H_+(x)\right]
&\geq
\dim E^u(x)+\dim H_+(x)-2n
\nonumber\\
&=
\dim E^u-n_{\mathrm{neg}} >0.
\label{E11}
\end{align}
Thus, $E^u(x)\cap H_+(x)\neq\{0\}$ for every finite $x$. Define $g_*=[\sup_{x\geq0}\|\mathbf{G}^{-1}(x)\|]^{-1}>0$. Since $\mathbf{G}(x)$ is Hermitian, every positive eigenvalue of $\mathbf{G}(x)$ is at least $g_*$, and therefore $w^{\dagger}\mathbf{G}(x)w\geq g_*\|w\|^2$ for every $w\in H_+(x)$.

Since $e^{-2\alpha x}\rightarrow0$, choose a sufficiently large but finite $X$ such that $\|\mathbf G(0)\|K^2e^{-2\alpha X}<g_*$, and take $0\neq w\in E^u(X)\cap H_+(X)$. Setting $v=U(0,X)w\in E^u$, Eq.~\eqref{app:backward-contraction} gives $\|v\|\leq Ke^{-\alpha X}\|w\|$. Monotonicity of the landscape along the forward trajectory from $v$ to $w$ then yields
\begin{equation}
\begin{aligned}
w^\dagger\mathbf G(X)w
&\leq v^\dagger\mathbf G(0)v\\
&\leq \|\mathbf G(0)\|\,\|v\|^2\\
&\leq \|\mathbf G(0)\|K^2e^{-2\alpha X}\|w\|^2\\
&<g_*\|w\|^2.
\end{aligned}
\end{equation}
However, $w\in H_+(X)$ requires $w^\dagger\mathbf G(X)w\geq g_*\|w\|^2$, giving a contradiction. Hence, $\dim E^u\leq n_{\mathrm{neg}}$. Finally, the direct-sum decomposition at $x=0$ gives
\begin{equation}
\dim E^s+\dim E^u=2n=n_{\mathrm{pos}}+n_{\mathrm{neg}}.
\end{equation}
Together with the above inequalities, this yields $\dim E^s=n_{\mathrm{pos}}$ and $\dim E^u=n_{\mathrm{neg}}$. This completes the proof. The initial direct-sum decomposition in Eq.~\eqref{app:anchored-splitting} is, for example, guaranteed by an exponential dichotomy
\cite{coppel1978dichotomies,sacker1978spectral}.

For a constant hyperbolic matrix $\mathbf M$, whose eigenvalues have nonzero real parts, $E^s$ and $E^u$ are generated by generalized eigenvectors associated with eigenvalues having negative and positive real parts, respectively. More generally, if $\mathbf{M}(x)$ approaches a hyperbolic constant matrix as $x\rightarrow+\infty$, the roughness theorem provides a sufficient condition for this splitting
\cite{coppel1978dichotomies,palmer1987perturbation}.

Not every stable solution of the zero-energy equation satisfies the physical open-boundary condition. Choose a basis of the stable subspace $\Phi_j^s=(\phi_j^s,\varphi_j^s)^T$, where $j=1,2,\ldots,n_{\mathrm{pos}}$. An arbitrary stable initial condition can be written as $\mathbf{\Phi}_0=\sum_{j=1}^{n_{\mathrm{pos}}}c_j\Phi_j^s$. The open-boundary condition requires the wave-function components to vanish, which is equivalent to $\mathbf A_s\mathbf c=0$, where the $j$th column of $\mathbf A_s$ is $\phi_j^s(0)$. Hence,
\begin{equation}
\mathcal{N}
=
\dim\ker\mathbf{A}_s
=
n_{\mathrm{pos}}
-\operatorname{rank}\mathbf{A}_s.
\end{equation}
This is Eq.~\eqref{eq11} of the main text.
For $n_{\mathrm{pos}}\geq n$, if the stable subspace
and the physical boundary subspace intersect transversely,
$\mathbf A_s$ has full row rank,
$\operatorname{rank}\mathbf A_s=n$.
The zero-mode count is then
\begin{equation}
\mathcal{N}
=
n_{\mathrm{pos}}-n
=
\frac{n_{\mathrm{pos}}-n_{\mathrm{neg}}}{2}.
\label{Aeq12}
\end{equation}

A complementary criterion excludes decaying boundary solutions. Suppose there exists a continuously differentiable, uniformly bounded Hermitian matrix $\mathbf G(x)$ such that $\mathbf\Gamma(x)\geq0$, and that $\mathcal L(0,\Phi_0)>0$ for every nonzero initial condition satisfying the physical boundary condition. Then $\mathcal L$ is nondecreasing along every spatial trajectory. If such an initial condition generated a trajectory tending to zero, boundedness of $\mathbf G(x)$ would imply $\mathcal L(x)\rightarrow0$. This contradicts $\mathcal L(x)\geq\mathcal L(0,\Phi_0)>0$. Therefore, no nonzero boundary initial condition generates a decaying solution; in the BdG setting considered here, this excludes a Majorana zero mode.

\textbf{Remark 1.}
The dimension identities in Theorem~1 remain valid for general
forward decay and growth rates, with all other conditions
retained. In place of Eq.~\eqref{app:anchored-splitting},
assume that $\mathbb C^{2n}=E^s\oplus E^u$ and that,
for some $K\geq1$,
\begin{equation}
\begin{aligned}
\|U(x,0)v_s\|
&\leq K\rho_s(x)\|v_s\|,
&&v_s\in E^s,\\
\|U(x,0)v_u\|
&\geq K^{-1}\rho_u(x)\|v_u\|,
&&v_u\in E^u,
\end{aligned}
\label{app:general-rates}
\end{equation}
where $\rho_s(x),\rho_u(x)>0$ satisfy
$\rho_s(x)\rightarrow0$ and $\rho_u(x)\rightarrow\infty$
as $x\rightarrow\infty$.
These bounds are uniform over vectors in the indicated
subspaces; the rate functions need not be exponential or monotone.

The stable-subspace argument is unchanged since
$\rho_s(x)\rightarrow0$.
For the unstable-subspace argument, choose a finite $X$ with
$\|\mathbf G(0)\|K^2\rho_u(X)^{-2}<g_*$.
For the nonzero vector $w\in E^u(X)\cap H_+(X)$
in the original contradiction argument, the same reasoning yields
$g_*\|w\|^2\leq w^\dagger\mathbf G(X)w<g_*\|w\|^2$. Thus, the dimension identities in Eq.~\eqref{A8} remain valid under these generalized bounds. 

To identify the resulting decaying boundary solutions with
normalizable zero modes, square integrability of the physical
wave functions must also be verified.
A sufficient condition is
$\int_0^\infty\rho_s(x)^2\,dx<\infty$.

\textbf{Remark 2.}
The existence of a uniformly bounded and uniformly invertible
landscape matrix satisfying $\mathbf\Gamma\leq0$ does not
by itself imply the splitting in
Eq.~\eqref{app:anchored-splitting}.
For example, $\mathbf M=\mathbf 0$ and $\mathbf G=\mathbf I$ give
$\mathbf\Gamma= \mathbf 0$, while every trajectory is constant.
Theorem~1 therefore assumes the decay and growth estimates
separately. The following strict Lyapunov criterion establishes
the required splitting~\cite{Mitropolsky2002}.

\textbf{Lemma 1.}
Let $\mathbf M(x)$ be continuous and uniformly bounded on
$[0,\infty)$. Suppose there exists a continuously differentiable Hermitian
matrix $\widetilde{\mathbf G}(x)$ with constant inertia
$\operatorname{In}[\widetilde{\mathbf G}(x)]=(p,q,0)$,
where $p+q=2n$.
Assume further that
$\|\widetilde{\mathbf G}(x)\|\leq g_1$ and
$\|\widetilde{\mathbf G}^{-1}(x)\|\leq g_0^{-1}$
hold for all $x\geq0$ and some constants $g_0,g_1>0$.
If, for some $c>0$,
\begin{equation}
\widetilde{\mathbf\Gamma}
:=
\partial_x\widetilde{\mathbf G}
+\widetilde{\mathbf G}\mathbf M
+\mathbf M^\dagger\widetilde{\mathbf G}
\leq -c \mathbf I
\label{app:strict-landscape}
\end{equation}
for all $x\geq0$, then the initial-value space admits a splitting
$\mathbb C^{2n}=E^s\oplus E^u$ with
$\dim E^s=p$ and $\dim E^u=q$.
The estimates in Eq.~\eqref{app:anchored-splitting} hold
for some constants $K\geq1$ and $\alpha>0$.

To establish the exponential splitting required by
Theorem~1 when the original landscape satisfies only
$\boldsymbol{\Gamma}(x)\leq0$, consider an auxiliary
landscape matrix of the form
$\widetilde{\mathbf G}(x)=\mathbf G(x)+\epsilon\mathbf A$,
where $\mathbf A$ is a constant Hermitian matrix and
$\epsilon>0$ is fixed. Its derivative matrix is  

\begin{equation}
\widetilde{\mathbf\Gamma}(x)
=
\mathbf\Gamma(x)
+\epsilon
\bigl[\mathbf A\mathbf M(x)+\mathbf M^\dagger(x)\mathbf A\bigr].
\end{equation}
The construction requires
$\widetilde{\mathbf\Gamma}(x)\leq-c\mathbf I$
uniformly for some $c>0$ and
$\delta:=\epsilon\|\mathbf A\|<g_*$, where
\begin{equation}
g_*=
\left[\sup_{x\geq0}\|\mathbf G^{-1}(x)\|\right]^{-1}
=
\inf_{x\geq0}\min_j|\lambda_j(x)|>0.
\end{equation}
Here, $\|\cdot\|$ denotes the spectral norm and
$\lambda_j(x)$ are the eigenvalues of $\mathbf G(x)$.
The Hermitian eigenvalue perturbation bound gives
\begin{equation}
|\widetilde{\lambda}_j(x)-\lambda_j(x)|
\leq\delta<g_*,
\end{equation}
with both sets of eigenvalues ordered increasingly.
Every eigenvalue therefore retains its sign and remains
at least $g_*-\delta>0$ away from zero. Thus,
$\operatorname{In}[\widetilde{\mathbf G}(x)]
=\operatorname{In}[\mathbf G(x)]$, and
\begin{equation}
\sup_{x\geq0}\|\widetilde{\mathbf G}^{-1}(x)\|
\leq\frac{1}{g_*-\delta}<\infty.
\end{equation}
Uniform boundedness follows from
$\|\widetilde{\mathbf G}(x)\|\leq\|\mathbf G(x)\|+\delta$.

If $\mathbf M(x)$ is also uniformly bounded, Lemma~1 establishes the exponential splitting in Eq.~\eqref{app:anchored-splitting}, with $\dim E^s=n_{\mathrm{pos}}$ and $\dim E^u=n_{\mathrm{neg}}$. These auxiliary conditions are verified for the parameter bounds in Eqs.~\eqref{eq6} and~\eqref{eq14} at the ends of Secs.~\ref{sec:sm-pwave} and~\ref{sec:sm-nanowire}, respectively, under the stated uniform bounds and positive margins. Theorem~1 therefore guarantees the existence of Majorana zero modes in both systems, completing the self-consistency verification of the two existence criteria.

\section{landscape construction for hybrid nanowires}
\label{sec:sm-nanowire}

Starting from Eq.~\eqref{nanowire_BdG}, the zero-energy equation for the Majorana mode localized at the left end can be recast as a four-component first-order spatial equation by combining the
two-component wave function and its spatial derivative into the
phase-space vector \(\mathbf{\Phi}\). We use the rescaled parameters
\(o_s\equiv 2mo/\hbar^2\), with
\(o\in\{B,\Delta,\mu,\alpha\}\), where the subscript \(s\) denotes the
rescaled quantities. The spatial evolution matrix reads
\begin{equation}
\mathbf M(x)=
\begin{bmatrix}
0&\mathbb{I}\\
B_s\sigma_z+i\Delta_s\sigma_y-\mu_s(x)I
&-i\alpha_s\sigma_y
\end{bmatrix}.
\end{equation}

For $B>\Delta>0$ and $\alpha>0$, we construct the quadratic
landscape
\begin{equation}
\mathcal L=\frac12\Phi^\dagger\mathbf G(x)\Phi,
\qquad
\mathbf G(x)=
\begin{bmatrix}
\mathrm a(x)&-\sigma_z\\
-\sigma_z&\mathrm d
\end{bmatrix},
\label{A13}
\end{equation}
where the upper-left block is
\begin{equation}
\mathrm a(x)=
\frac{B_s\mu_s(x)}{\alpha_s\Delta_s}\mathbb{I}
+\left(\frac{\mu_s(x)}{\alpha_s}-\alpha_s\right)\sigma_x
-\frac{\kappa_s^2}{\alpha_s\Delta_s}\sigma_z,
\end{equation}
while the lower-right block is
$\mathrm d=[(B_s /\Delta_s)\mathbb{I} +\sigma_x]/\alpha_s$, with
$\kappa_s=\sqrt{B_s^2-\Delta_s^2}$.
The spatial-evolution equation gives
$\partial_x\mathcal L=(1/2)\Phi^\dagger
\boldsymbol{\Gamma}(x)\Phi$, where
\begin{equation}
\boldsymbol{\Gamma}(x)=
\begin{bmatrix}
2\,\mathrm m(x)+\partial_x\mathrm a(x)&0\\
0&0
\end{bmatrix},
\end{equation}
with $\mathrm m(x)=-B_s\mathbb{I}-\Delta_s\sigma_x+\mu_s(x)\sigma_z$.

For the landscape constructed above,
$|\mu_s(x)|<\kappa_s$ ensures
$\operatorname{In}[\mathbf G(x)]=(3,1,0)$. Meanwhile, the monotonicity condition
$\boldsymbol{\Gamma}(x)\leq0$ yields
\begin{equation}
\sqrt{\left(\Delta-\frac{1}{2\alpha_s}\partial_x\mu\right)^2
+\mu^2}
<
B-\frac{B}{2 \alpha_s \Delta}\partial_x\mu.
\end{equation}
Under the conditions of Theorem~1,
Eq.~\eqref{eq11} gives
$\mathcal N=3-\operatorname{rank}\mathbf A_s\geq1$.
If the transversality condition for Eq.~\eqref{Aeq12}
also holds, $\mathcal N=1$, corresponding to one Majorana
zero mode localized at the left end.
The exponential splitting required by Theorem~1
is verified at the end of this section under
Eq.~\eqref{eq14} with a uniform positive margin.

The zero-energy equation for the Majorana mode localized at the right end can be analyzed in the same way by reversing the spatial coordinate. Under this transformation, \(\partial_x\mu\) changes sign,
while the local condition on \(\mu\) remains unchanged. The corresponding gradient condition is obtained by replacing \(\partial_x\mu\rightarrow-\partial_x\mu\). Therefore, a sufficient
condition for the existence of Majorana zero modes at both ends is
\(\left| \mu(x) \right| < \sqrt{B^{2} - \Delta^{2}}\) and
\begin{equation}
\left| \partial_{x}\mu(x) \right| <
\frac{4\Delta}{l_{\mathrm{so}}}
\left(
1-\frac{\left| \mu(x) \right|}
{\sqrt{B^{2}-\Delta^{2}}}
\right). 
\end{equation}
where $l_{\mathrm{so}}=\hbar^2/(m\alpha)$. This is Eq.~\eqref{eq14} of the main text.

Finally, we self-consistently verify that the parameter bounds
given in Eq.~\eqref{eq14} ensure that the conditions
for applying Theorem~1 are satisfied.
Assume $\mu\in C^1([0,\infty))$ and impose
Eq.~\eqref{eq14} with a uniform positive margin:
\begin{equation}
\frac{|\mu_s(x)|}{\kappa_s}
+\frac{|\mu_s'(x)|}{2\alpha_s\Delta_s}
\leq1-\eta,
\qquad 0<\eta<1.
\label{app:nanowire-uniform}
\end{equation}
Introduce $\chi(x)=1-\mu_s'(x)/(2\alpha_s\Delta_s)$
and write the nonzero block of $\boldsymbol{\Gamma}$
as $-\mathbf R(x)$. In terms of the blocks defined above,
\begin{equation}
\begin{aligned}
\mathbf R(x)
&=-2\,\mathrm m(x)-\partial_x\mathrm a(x)\\
&=2\chi(x)(B_s\mathbb I+\Delta_s\sigma_x)
-2\mu_s(x)\sigma_z.
\end{aligned}
\end{equation}
Equation~\eqref{app:nanowire-uniform} implies
$\chi-\eta\geq|\mu_s|/\kappa_s\geq0$.
The triangle inequality and
$B_s^2=\Delta_s^2+\kappa_s^2$ therefore give
\begin{equation}
\begin{aligned}
\sqrt{\Delta_s^2\chi^2+\mu_s^2}
&\leq\Delta_s\eta
+\sqrt{\Delta_s^2(\chi-\eta)^2+\mu_s^2}\\
&\leq\Delta_s\eta+B_s(\chi-\eta).
\end{aligned}
\end{equation}
Consequently,
\begin{equation}
\begin{aligned}
\lambda_{\min}[\mathbf R(x)]
&=2\left[B_s\chi-\sqrt{\Delta_s^2\chi^2+\mu_s^2}\right]\\
&\geq r_0:=2(B_s-\Delta_s)\eta>0.
\end{aligned}
\end{equation}
Thus, $\mathbf R(x)\geq r_0\mathbb I$ uniformly,
providing a uniform negative margin in the nonzero
block of $\boldsymbol{\Gamma}$.

The amplitude part of the same constraint confines $\mu_s$
to the compact interval $|\mu_s|\leq(1-\eta)\kappa_s$.
Since $\mathbf G$ depends continuously on $\mu_s$ and
has inertia $(3,1,0)$ throughout this interval,
it is uniformly bounded and uniformly invertible.
Let $g_*>0$ denote a common lower bound on the absolute
values of its eigenvalues over this interval.
The same coefficient bounds ensure that $\mathbf M$
is uniformly bounded.

With these uniform bounds established, consider the auxiliary
landscape matrix
$\widetilde{\mathbf G}(x)
=\mathbf G(x)-\epsilon(\sigma_x\otimes\mathbb I)$.
The corresponding derivative matrix is
\begin{equation}
\widetilde{\boldsymbol{\Gamma}}(x)
=
\begin{bmatrix}
-\mathbf R-2\epsilon(B_s\sigma_z-\mu_s\mathbb I)
&i\epsilon\alpha_s\sigma_y\\
-i\epsilon\alpha_s\sigma_y
&-2\epsilon\mathbb I
\end{bmatrix}.
\end{equation}
Define $\beta_s=2[B_s+(1-\eta)\kappa_s]$, so that
$\|2(B_s\sigma_z-\mu_s\mathbb I)\|\leq\beta_s$
uniformly.

For any phase-space vector
$\mathbf\Phi=(\boldsymbol\phi,\boldsymbol\varphi)^T$,
with $\boldsymbol\phi,\boldsymbol\varphi\in\mathbb C^2$,
Young's inequality
$2\alpha_s\|\boldsymbol\phi\|\|\boldsymbol\varphi\|
\leq\alpha_s^2\|\boldsymbol\phi\|^2
+\|\boldsymbol\varphi\|^2$
and the preceding matrix bounds yield
\begin{equation}
\mathbf\Phi^\dagger
\widetilde{\boldsymbol{\Gamma}}\mathbf\Phi
\leq-
\left[r_0-\epsilon(\beta_s+\alpha_s^2)\right]
\|\boldsymbol\phi\|^2 -\epsilon\|\boldsymbol\varphi\|^2.
\end{equation}
Choose a fixed
$0<\epsilon<\min\{g_*,r_0/(\beta_s+\alpha_s^2)\}$.
The derivative estimate then gives
$\widetilde{\boldsymbol{\Gamma}}(x)\leq-c_n\mathbb I_4$,
where
$c_n=\min\{\epsilon,r_0-\epsilon(\beta_s+\alpha_s^2)\}>0$.

For the same fixed $\epsilon$,
$\|\widetilde{\mathbf G}(x)-\mathbf G(x)\|=\epsilon<g_*$.
The Hermitian eigenvalue perturbation bound ensures that
$\widetilde{\mathbf G}(x)$ retains the inertia $(3,1,0)$,
with all eigenvalues remaining at least $g_*-\epsilon>0$
away from zero. Consequently,
\begin{equation}
\sup_{x\geq0}\|\widetilde{\mathbf G}^{-1}(x)\|
\leq\frac{1}{g_*-\epsilon}<\infty.
\end{equation}
Uniform boundedness is also preserved, since
$\|\widetilde{\mathbf G}(x)\|
\leq\|\mathbf G(x)\|+\epsilon$
and $\mathbf G$ is uniformly bounded.

All conditions of Lemma~1 in Sec.~\ref{sec:sm-theorem} are therefore satisfied.
The lemma establishes the exponential splitting in
Eq.~\eqref{app:anchored-splitting}, with
$\dim E^s=3$ and $\dim E^u=1$.
Thus, the amplitude--gradient bound, imposed with the stated
uniform positive margin, ensures the dynamical condition
required by Theorem~1.
The same verification applies to the right-boundary branch,
since Eq.~\eqref{app:nanowire-uniform} is invariant under
reversal of the spatial coordinate.

\subsection{Numerical verification of the profile-independent criterion}
\label{sec:sm-numerics}

In this subsection, we numerically demonstrate the
profile-independent predictions of Eq.~\eqref{eq14}
for three distinct potentials sharing the same amplitude
and gradient bounds.

The numerical calculations in Fig.~\ref{Fig4} use the
low-energy Hamiltonian of a nanowire--superconductor hybrid
system in Eq.~\eqref{nanowire_BdG}.
The parameters are taken to be $B=0.8\,\mathrm{meV}$,
$\Delta=0.2\,\mathrm{meV}$, $l_{\mathrm{so}}=100\,\mathrm{nm}$,
and $m=0.023m_e$.
These values give the spin-orbit coupling
$\alpha=\hbar^2/(m l_{\mathrm{so}})
\simeq33.1303\,\mathrm{meV\,nm}$ and
$\kappa=\sqrt{B^2-\Delta^2}\simeq0.774597\,\mathrm{meV}$.

The three chemical-potential profiles are
$\mu_j(x)=Af_j(q_jx)$, with $q_j=g/(AC_j)$.
Here, $A=\kappa/2$ and $g=3\,\mu\mathrm{eV/nm}$
specify common bounds on the chemical-potential magnitude
and spatial gradient, respectively.
The dimensionless functions are
\begin{equation}
\begin{aligned}
f_1(s)&=\cos s,\\
f_2(s)&=\frac{2\cos s+\cos 2s}{3},\\
f_3(s)&=\frac{\cos s+\cos(\sqrt{2}s)}{2},
\end{aligned}
\label{eq:sm-profile-forms}
\end{equation}
where $C_1=1$, $C_2=4/3$, and $C_3=(1+\sqrt{2})/2$
satisfy $|f_j'(s)|\leq C_j$.
No phase shifts or constant offset are included.
Since $|f_j|\leq1$, the chosen wave numbers ensure
$|\mu_j(x)|\leq A$ and $|\partial_x\mu_j(x)|\leq g$
for all $x\geq0$.
The gradient bound $g$ need not equal the maximum gradient
of each profile.
The first two profiles are periodic and the third is
quasiperiodic; all remain spatially modulated at
arbitrarily large distances.

The common bounds satisfy
\begin{equation}
\frac{A}{\kappa}+\frac{l_{\mathrm{so}}g}{4\Delta}
=\frac12+\frac38=\frac78<1,
\label{eq:sm-common-certificate}
\end{equation}
so Eq.~\eqref{eq14} holds with a uniform positive margin
$\eta=1/8$.
The self-consistency verification above ensures that
Theorem~1 applies to every continuously differentiable
profile obeying these common bounds and guarantees
the existence of Majorana zero modes.
The finite-wire calculations below examine the predicted
end localization and finite-size energy splitting
for the three explicit profiles.

For the finite-wire calculations, Dirichlet boundary
conditions are imposed at $x=0$ and $x=L$.
The interior sites are $x_i=ih$, with $i=1,\ldots,N$
and $N=L/h-1$.
A centered finite-difference discretization gives the
onsite and forward-hopping blocks
\begin{equation}
\begin{aligned}
H_{ii}
&=(2t-\mu_i)\mathbb I\otimes\tau_z
+B\sigma_z\otimes\tau_z-\Delta\sigma_y\otimes\tau_y,\\
H_{i,i+1}
&=-t\mathbb I\otimes\tau_z
-\frac{i\alpha}{2h}\sigma_y\otimes\tau_z,
\end{aligned}
\label{eq:sm-numerical-bdg}
\end{equation}
where $t=\hbar^2/(2mh^2)$, $\mu_i=\mu_j(x_i)$,
and $H_{i+1,i}=H_{i,i+1}^\dagger$.
The first and second tensor factors refer to spin and
particle--hole space, respectively.

The calculations use $h=2.5\,\mathrm{nm}$ and wire lengths
from $2$ to $6\,\mu\mathrm m$ in $50\,\mathrm{nm}$ steps.
Sparse shift-invert diagonalization computes six
eigenpairs near zero, which are then sorted by $|E|$. The plotted quantity is
$\varepsilon(L)=(|E_0|+|E_1|)/2$, where $E_0$ and $E_1$
are the two eigenvalues closest to zero.
Their energy separation is approximately $2\varepsilon(L)$.
The corresponding eigenstates are linearly combined
to form states localized at the left and right ends.
These combinations minimize and maximize the mean
position within the two-state subspace, respectively.
The inset shows their probability densities, each
rescaled to unit peak. The eigenvalues and eigenstates are computed directly
from the discretized BdG Hamiltonian, without using
the landscape or auxiliary matrices.

Spatial discretization is checked by recomputing all
three profiles at $L=2$, $4$, and $6\,\mu\mathrm m$
using $h=5$ and $1.25\,\mathrm{nm}$.
The relative difference in $\varepsilon(L)$ between
the $h=2.5$ and $1.25\,\mathrm{nm}$ grids remains below
$0.46\%$, using the $h=2.5\,\mathrm{nm}$ value as the
reference.
Across all calculations, the residual norm
$\|H\psi-E\psi\|$ of the eigenvalue equation is below
$1.1\times10^{-9}\,\mathrm{meV}$ for every normalized
eigenstate.
For the two eigenstates closest to zero, the residuals
are below $2.3\times10^{-10}\,\mathrm{meV}$, and their
largest ratio to $\varepsilon(L)$ is below
$1.4\times10^{-3}$.
These checks assess the numerical accuracy of the
finite-difference calculation.
Finite wires generally exhibit a nonzero, oscillatory
splitting due to overlap between the end states;
the analytical criterion guarantees exact zero modes
in the semi-infinite limit.

For the comparison in Fig.~\ref{Fig4}(c), the infinite
uniform continuum system with $\mu=0$ is taken as the
reference. Its lower positive-energy band satisfies
\begin{equation}
\begin{aligned}
\bigl[E_-(k)\bigr]^2
&=\xi_k^2+\alpha^2k^2+B^2+\Delta^2\\
&\quad-2\sqrt{B^2(\xi_k^2+\Delta^2)
+\xi_k^2\alpha^2k^2},
\end{aligned}
\label{eq:sm-reference-dispersion}
\end{equation}
where $\xi_k=\hbar^2k^2/(2m)$.
Minimizing this band gives
$E_g\simeq0.148384\,\mathrm{meV}$ at
$|k|\simeq0.0266725\,\mathrm{nm}^{-1}$.
This value is checked against direct diagonalization
of the uniform $4\times4$ Bloch Hamiltonian at the
minimizing wave vector.

The spatially varying chemical potential adds
$V(x)=-\mu(x)\mathbb I\otimes\tau_z$ to the reference
Hamiltonian $H_0$, with operator norm $\|V\|\leq A$.
The condition $A<E_g$ therefore ensures that the bulk
gap remains open along $H_0+\lambda V$,
$0\leq\lambda\leq1$, without imposing a gradient
constraint.
The chosen bound $A\simeq0.387298\,\mathrm{meV}$
exceeds $E_g$, whereas the common amplitude and
gradient bounds satisfy
Eq.~\eqref{eq:sm-common-certificate}.
The landscape criterion therefore certifies these
profiles beyond the range covered by this particular
norm estimate.

\subsection{Analytical bounds for a Gaussian modulation}
\label{sec:sm-analysis}
\begin{figure}
\includegraphics[width=6cm]{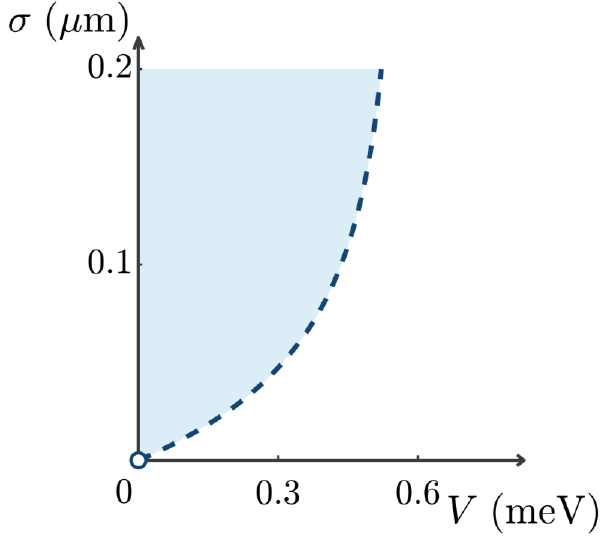}
\caption{Sufficient amplitude–width region for the Gaussian modulation in Eq. (S46). The blue-shaded interior above the dashed curve satisfies Eq. (14). The dashed curve marks equality in Eq. (S49) and is excluded by the strict criterion. Parameters in \(\Delta=0.2\) meV, \(l_{\mathrm{so}}=100\) nm, \(\mu_0=0.2\) meV, and \(B=0.8\) meV.}
\label{SFig1}
\end{figure}

We now use Eq.~\eqref{eq14} to derive explicit sufficient
bounds on the strength and width of a Gaussian modulation.
For a localized impurity or gate-induced potential~\cite{Brouwer2012,Pan2020},
we consider the profile
\begin{equation}
\mu(x)=\mu_0+V\exp\left[-\frac{(x-x_0)^2}{2\sigma^2}\right].
\end{equation}
Here, $\mu_0>0$ is the background chemical potential, and $V>0$, $x_0$, and $\sigma$ denote the modulation amplitude, center, and width, respectively. The local condition
\(\mu(x)<\kappa\) gives the amplitude bound
\(V<\kappa-\mu_{0}\), where \(\kappa=\sqrt{B^{2}-\Delta^{2}}\). To impose the gradient condition in Eq.~\eqref{eq14} throughout the
wire, introduce the dimensionless variable
\(t=(x-x_{0})^{2}/(2\sigma^{2})\geq0\). For the Gaussian profile, one has
\(\left|\partial_x\mu\right|=V\sqrt{2t}\,e^{-t}/\sigma\).
Using \(\mu(x)=\mu_0+Ve^{-t}\), the gradient condition in
Eq.~\eqref{eq14} becomes
\begin{equation}
\frac{1}{\sigma}<\frac{4\Delta}{l_{\mathrm{so} }V}
\frac{e^{t}(1-\mu_{0}/\kappa)-V/\kappa}{\sqrt{2t}}.
\label{A20}
\end{equation}
Since Eq.~\eqref{A20} must hold for every $t>0$,
the upper bound on $1/\sigma$ is determined by the minimum of
$f(t)=[e^t(1-\mu_0/\kappa)-V/\kappa]/\sqrt{2t}$. For $0<V<\kappa-\mu_0$, this function diverges as
$t\rightarrow0^+$ and $t\rightarrow\infty$.
Its unique minimum occurs at $t=t_m$, where
\begin{equation}
e^{t_m}(2t_m-1)
=-\frac{V}{\kappa-\mu_0},
\qquad 0<t_m<\frac12.
\end{equation}
For $t>0$, $e^t(2t-1)$ is strictly increasing,
so this stationary point is unique. Using this stationarity condition gives
$f(t_m)=(1-\mu_0/\kappa)e^{t_m}\sqrt{2t_m}$.
Equation~\eqref{A20} consequently becomes
\begin{equation}
\frac{1}{\sigma}
<
\frac{4\Delta}{l_{\mathrm{so}}V}
\left(1-\frac{\mu_0}{\kappa}\right)
e^{t_m}\sqrt{2t_m}.
\label{A24}
\end{equation}
Together with $0<V<\kappa-\mu_0$, this inequality
defines the sufficient region in the
$(V,\sigma)$ plane shown in
Fig.~\ref{SFig1}.

These strict amplitude and width bounds ensure a uniform
positive margin in Eq.~\eqref{eq14}: its left-hand side
is continuous and approaches $\mu_0/\kappa<1$ far from
the modulation. The self-consistency verification above
therefore applies, and Theorem~1 guarantees the existence
of Majorana zero modes. This example translates the
general amplitude--gradient criterion into explicit
sufficient constraints on the modulation strength
and width.

\end{document}